\documentclass[floatfix,amsmath,amssymb,nofootinbib,superscriptaddress,longbibliography,twocolumn, prx]{revtex4-2}
	\usepackage{chngcntr}
	\usepackage{graphicx}
	\usepackage{color}
	\usepackage{amsmath}
	\usepackage{enumitem}
	\usepackage{amssymb}
	\usepackage{cancel}
	\usepackage{ulem}
	\usepackage{multirow}
    \usepackage[colorlinks=true,linkcolor=blue,citecolor=blue]{hyperref}

	\newcommand{\be}{\begin{equation}}
		\newcommand{\ee}{\end{equation}}
	
	\newcommand{\bea}{\begin{eqnarray}}
		\newcommand{\eea}{\end{eqnarray}}

	\newcommand{\lb}{\left[}
	\newcommand{\rb}{\right]}
	\newcommand{\lp}{\left(}
	\newcommand{\rp}{\right)}
	
	\newcommand{\Tr}{{\rm \, Tr\,}}
	
	\renewcommand{\Im}{{\rm \, Im\,}}

	\renewcommand{\vec}[1]{{\boldsymbol #1}}

\usepackage[dvipsnames]{xcolor}
\newcommand{\approve}[1]{\textcolor{black}{#1}}

\newcommand{\addELH}[1]{\textcolor{black}{#1}} 

\usepackage{bbold}

\usepackage{orcidlink}
 
\begin{document}
		\title{Superconductivity from spin-canting fluctuations in rhombohedral graphene}

\author{Zhiyu Dong\,\orcidlink{0000-0003-3979-914X}}
\affiliation{Department of Physics and Institute for Quantum Information and Matter, California Institute of Technology, Pasadena, CA 91125, USA}

\author{\'Etienne Lantagne-Hurtubise\,\orcidlink{0000-0003-0417-64521}}
\affiliation{Department of Physics and Institute for Quantum Information and Matter, California Institute of Technology, Pasadena, CA 91125, USA}
\affiliation{Département de physique and Institut quantique, Université de Sherbrooke, Sherbrooke, Québec J1K 2R1, Canada}

\author{Jason Alicea\,\orcidlink{0000-0001-9979-3423}}
\affiliation{Department of Physics and Institute for Quantum Information and Matter, California Institute of Technology, Pasadena, CA 91125, USA}

		\begin{abstract}
  Rhombohedral graphene multilayers host various broken-symmetry metallic phases as well as superconductors whose pairing mechanism and order parameter symmetry remain unsettled. Strikingly, experiments have revealed prominent new superconducting regions in rhombohedral bilayer and trilayer graphene devices with proximity-induced Ising spin-orbit coupling. We propose that these superconductors descend from a common spin-canted normal state that spontaneously breaks a U(1) spin symmetry and thus supports soft magnon modes. In particular, we show that these soft modes can mediate pairing through inter-band scattering events that are symmetry-forbidden in the absence of spin-orbit coupling, thus providing a promising explanation for spin-orbit-enabled pairing. Numerous other experimental observations---including nontrivial dependence of superconductivity on the spin-orbit coupling strength, in-plane magnetic fields, and Fermi surface structure---also naturally follow from our scenario.   
        \end{abstract}
		\maketitle
		
\section{Introduction}

Rhombohedral graphene---composed of layers stacked in an `ABC' pattern---provides an extremely clean, gate-tunable platform for emergent quantum phenomena~\cite{Weitz2010, Shi2020, zhou2021half, zhou2021superconductivity, zhou2022isospin, Seiler2022, delaBarrera2022,  Kerelsky2021, Han2023, liu2023interactiondriven, Han2023a}.  As prominent examples, `pure' bilayer and trilayer devices (i.e., without proximity effects from adjacent 2D materials) exhibit exceptionally rich phase diagrams. Indeed, electrically varying the density and perpendicular displacement field reveals a series of correlated metallic states
as well as unconventional superconductors~\cite{zhou2021superconductivity, zhou2022isospin, Seiler2022, delaBarrera2022}.  Despite detailed distinctions
such as fermiology, normal-state order, and response to magnetic fields, 
these superconducting states share a common unifying ingredient:
Each one appears tethered to a narrow density window close to a normal-state phase transition, prompting the investigation of
pairing scenarios based on critical fluctuations~\cite{Chatterjee2022,dong2023signatures,Dong2023,dong2023superconductivity} and other mechanisms~\cite{Berg2021, chou2021acoustic,Chou2022,szabo2022competing,Szab2022,You2022,cea2022superconductivity,Lu2022,cea2023superconductivity,qin2023functional, Li2023, Dai2023, Jimeno-Pozo2022, Wagner2024,son2024,Chau2024originsuperconductivityrhombohedraltrilayer,Rubio2024}. 

Placing rhombohedral bilayers and trilayers proximate to a transition metal dichalcogenide (TMD) effectively boosts graphene's spin-orbit coupling by orders of magnitude~\cite{Gmitra2016, Wang2016, Gmitra2017, Khoo2017, Wang2019, Island2019, Amann2022, Zhang2023, Holleis2025, Yiran2025, li2024tunable, Patterson2025, LongJu2025, Sha2024, Han2024}, in turn
enriching the phenomenology in a very interesting way.  Namely, new superconducting phases emerge that display the following salient features~\cite{Zhang2023, Holleis2025, Patterson2025, li2024tunable, Yiran2025, LongJu2025}: $(i)$ Rather than residing near a phase boundary, they occur over a broad density range within a normal state that hosts two large
majority Fermi pockets coexisting with some number of small 
minority pockets. $(ii)$ Their optimal critical temperatures $T_c$ far exceed those arising in `pure' devices.
And $(iii)$ their susceptibility to depairing by in-plane magnetic fields varies nontrivially across the superconducting regions.  Numerous scenarios for how spin-orbit coupling promotes Cooper pairing in this setting have been introduced~\cite{Yangzhi2022,Curtis2023,Jimeno-Pozo2022,Wagner2024,Ming2023,Shavit2023,son2024}. 
Recent experiments on bilayers~\cite{Yiran2025} further showed that, as the spin-orbit coupling imparted by the TMD increases, the strongest new superconductor 
populates a diminished area in the phase diagram---yet curiously exhibits an enhanced maximal $T_c$.   

Our goal here is to develop a unified scenario that accounts for these characteristics of (ostensibly) spin-orbit-enabled superconductors in both bilayers and trilayers.  More specifically, we will address the following questions: What is the appropriate normal state that gives way to 
superconductivity?  What is the pairing mechanism, and how is it `activated' by spin-orbit coupling?  Given a putative normal state and pairing mechanism, how can one recover the detailed phenomenology highlighted above (which places
stringent constraints on theory)?

It is instructive to first address the fate of the normal state in `pure' bilayer and trilayer devices \emph{without} an adjacent TMD, where to a good approximation the system preserves SU(2) spin-rotation symmetry.  In the relevant density and displacement field regime, quantum oscillation measurements clearly evidence lifting of graphene's spin/valley degeneracy by interactions, yielding a broken-symmetry metallic state with majority and minority carriers.
We will assume that this phase corresponds to a simple Stoner ferromagnet that spontaneously breaks the SU(2) symmetry by developing an arbitrarily oriented spin magnetization.  This assumption is bolstered by experiments (specifically, magnetic field dependence of the phase boundary with adjacent symmetric and/or spin-and-valley polarized states, as well as a vanishing anomalous Hall signal~\cite{zhou2021half, zhou2022isospin}). Moreover, Hartree-Fock simulations find that this order is stabilized by long-range Coulomb repulsion supplemented by ferromagnetic Hund's coupling $J$ that favors aligning the spins in the $K$ and $K'$ valleys~\cite{zhou2021half, Szab2022, Chatterjee2022, Arp2024, Koh2024, Koh2024a}.  

\begin{figure}
    \centering
    \includegraphics[width = 0.95\columnwidth]{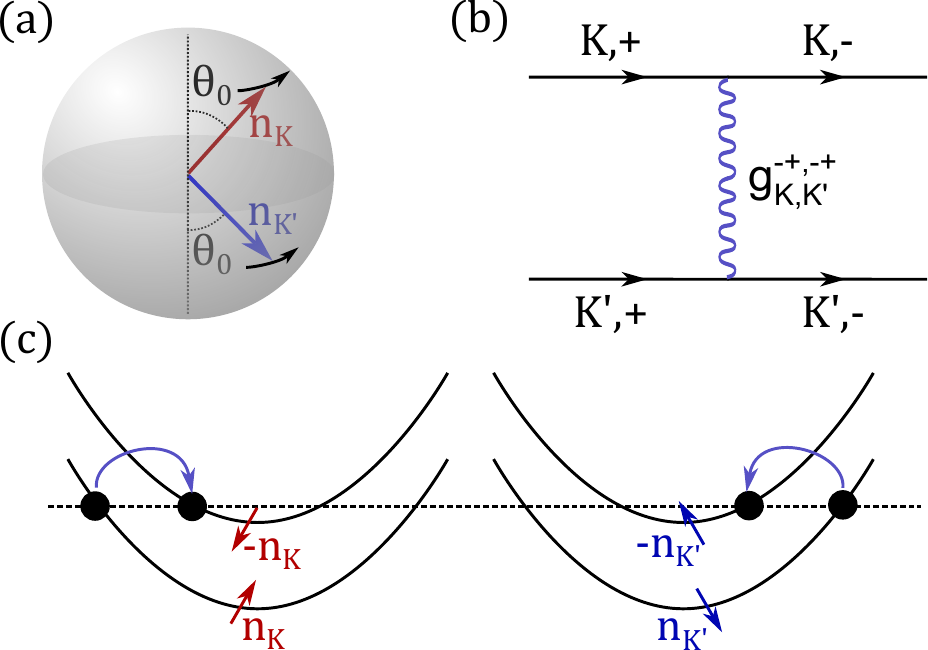}
        \caption{\textbf{Summary of the proposed pairing mechanism.} (a) Illustration of spin-canting order. The spin polarization $\vec{n}_{K}$ ($\vec{n}_{K'}$) in each valley is represented by a red (blue) arrow on the Bloch sphere.  Ising spin-orbit interaction defines an out-of-plane quantization axis that points in opposite directions in the two valleys. Spin-canted order arises when an in-plane magnetic moment is spontaneously generated, leading to a non-trivial polar angle $\theta_0$ of the magnetization vectors
        and an arbitrarily chosen azimuthal angle. Gapless magnon (Goldstone) modes associated with the spontaneous symmetry breaking are shown by black arrows. Soft magnon modes can mediate pairing through interband (spin-flipping) scattering, depicted with (b) Feynman diagrams and (c) kinematics.}
        \label{fig1}
\end{figure}

Anchored by this plausible ansatz, we then ask how the Stoner ferromagnet evolves upon resurrecting spin-orbit interaction inherited from the TMD.  Throughout this work we 
assume that the TMD primarily engenders Ising spin-orbit coupling $\lambda_I$, which preserves a U(1) subgroup of SU(2) corresponding to spin rotations about an out-of-plane 
Ising axis.  By itself, $\lambda_I$ favors orienting the spins normal to the graphene layers, in opposite directions for the two valleys---clearly competing with the 
different spin profile favored by Hund's coupling. 
As demonstrated by Hartree-Fock simulations~\cite{Koh2024, Koh2024a}, over a finite window of $\lambda_I$ the system can reach a compromise by realizing a spin-canted phase:  Here, the spins tilt away from the Ising axis by a canting angle $\theta_0$ as shown in Fig.~\ref{fig1}, spontaneously breaking the U(1) symmetry by developing an in-plane spin magnetization that partially satisfies $J$. This logic leads us to postulate that \emph{the normal state for spin-orbit-enabled superconductors in both bilayers and trilayers hosts spin-canting order}.  

Given that the spin-canted phase evolves smoothly from a simple Stoner ferromagnet upon turning on $\lambda_I$, one can then ask why the former would be more conducive to Cooper pairing than the latter.  A first hint surfaces from examining their collective modes.  Both phases spontaneously break a continuous symmetry---U(1) spin rotations for the spin-canted phase and SU(2) for the 
Stoner ferromagnet---and hence both support 
soft order parameter fluctuations: a gapless Goldstone magnon excitation, and a spin-canting mode with parametrically small gap that becomes critical at the canting transition.
A qualitative distinction nevertheless exists. Namely, the simple Stoner ferromagnet preserves a U(1) subgroup of SU(2) spin rotations, and hence the  magnetization must be preserved by arbitrary magnon-mediated processes. Might then soft magnon modes provide a pairing glue whose effectiveness depends critically on the symmetry-allowed scattering events that they mediate?

In the case with spin-orbit coupling, we indeed show that magnons that are softened near or inside the spin-canted phase can mediate attraction through spin-flipping, interband scattering that
favors $s$-wave pairing with an opposite sign between the majority and minority Fermi surfaces\footnote{Throughout this paper, we refer to a ``pseudospin-singlet'' $s$-wave channel where the pairing gap $\Delta({\bf p})$ has no sign alternation when ${\bf p}$ moves along each Fermi surface. However, due to fermion anticommutation relation, $\Delta({\bf p})$ changes sign when $\bf p$ moves between valleys. When viewed in the whole hexagonal Brillouin zone, this pairing channel can thus be classified as $f$-wave.\label{footnote_swave}.
The energy of the soft magnon modes (over the relevant momentum range for inter-band scattering) constitutes a natural low-frequency scale, providing retardation effects that are crucial to overcome the pair-breaking tendencies of Coulomb repulsion. In the absence of spin-orbit coupling, by contrast,
magnons of the SU(2) Stoner ferromagnet yield a vanishing inter-band pairing interaction due to constraints imposed by the remaining U(1) symmetry, thus furnishing a plausible explanation for spin-orbit-enabled superconductivity in rhombohedral graphene.} 

Our 
scenario also accommodates more refined experimental observations enumerated earlier. First, recall that spin-orbit-enabled superconductivity entails
minority Fermi pockets; moreover, in trilayers superconductivity abruptly turns off when they are vacated through a Lifshitz transition~\cite{Patterson2025}. These observations hint that pairing requires the presence of minority carriers, in harmony with spin-flipping scattering that acts between the majority and minority Fermi surfaces.
Second, Ref.~\onlinecite{Yiran2025} observed a tunable and gradual increase of the maximal $T_c$ with induced Ising spin-orbit-coupling $\lambda_I$, coinciding with a \addELH{\emph{reduced} (yet still extended)} phase space occupied by superconductivity. The concurrence of these two trends is surprising because in a conventional scenario, where the pairing glue is non-magnetic (e.g., phonon-mediated), spin-orbit coupling is expected to either uniformly help or hurt superconductivity through its effect on the Fermi surface. Our scenario, however, naturally recovers these seemingly at odds trends: Increasing $\lambda_I$ decreases the canting angle $\theta_0$ [see Eq.~\eqref{canting_angle}]---in turn enhancing the magnon-mediated pairing interaction [see Eq.~\eqref{eq:g at theta=0}] and thereby boosting $T_c$.
At the same time, spin-orbit interaction competes against Hund's coupling, such that increasing $\lambda_I$ erodes the phase space occupied by the spin-canted phase, whose Goldstone modes provide the pairing glue in our scenario. 
Increasing $\lambda_I$ thus also narrows the superconducting phase. 
Recent Hartree-Fock simulations support the predicted change in $\theta_0$ with $\lambda_I$ \cite{Koh2024a} accounting for the rise in $T_c$, while also finding that the phase space for spin-canting order evolves with $\lambda_I$ similarly to the phase space for superconductivity 
in experiment (see Fig.~7 in Ref.~\onlinecite{Koh2024} and Fig.~4 in Ref.~\onlinecite{Koh2024a}).
To the best of our knowledge, no other pairing mechanism relates superconductivity to spin canting in these systems.

Our theory may also explain unusual trends seen in the effects of in-plane magnetic fields on spin-orbit-enabled superconductivity. While naively one might anticipate high tolerance to in-plane magnetic fields due to Ising spin-orbit interaction~\cite{Lu2015, Xi2015}, experiments have measured critical in-plane fields that vary from well above the Pauli limit to below the Pauli limit depending on parameters~\cite{Zhang2023, Holleis2025, li2024tunable, Patterson2025, Yiran2025}.  Orbital or Rashba spin-orbit effects may at least partially explain these trends, though our scenario predicts a new depairing mechanism that could contribute to the in-plane field response. 
Specifically, in-plane magnetic fields explicitly break the U(1) spin symmetry that the spin-canted phase breaks spontaneously---thereby gapping out the Goldstone mode and reducing the effectiveness of magnon-mediated pairing. This behavior stands in contrast to non-magnetic pairing glues, such as phonons, where superconductivity would be suppressed primarily by field-induced breaking of pairing resonances at the Fermi energy.

\begin{figure*}
    \centering
    \includegraphics[width = 1.8\columnwidth]{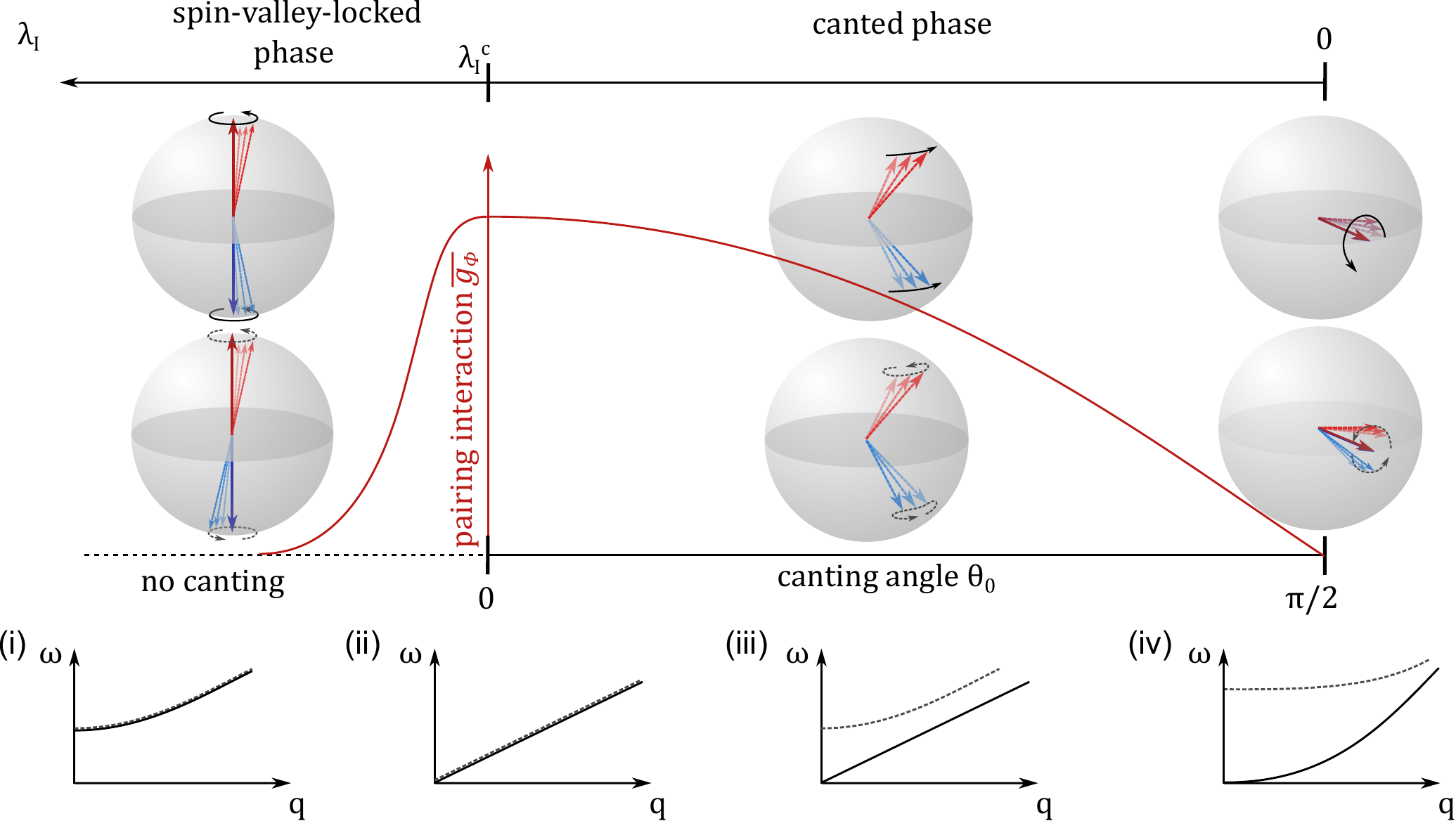}
    \caption{ \textbf{Magnon modes and pairing interaction.} The upper line illustrates the presumed phase diagram as a function of Ising spin-orbit coupling $\lambda_I$; a spin-canted phase appears up to a critical value $\lambda_I^c$, beyond which a symmetric spin-valley-locked phase emerges. 
    The red curve sketches the magnon-mediated pairing interaction strength [see Eq.~\eqref{eq:g_eff}], which depends sensitively on the canting angle $\theta_0$ and hence on spin-orbit coupling.
    Spheres illustrate the two relevant low-energy magnon modes in each regime: a valley-symmetric mode ($s-$, solid black arrow) and a valley-antisymmetric mode ($a-$, dashed gray arrow)---in valleys $K$ (blue arrows) and $K'$ (red arrows). Insets (i)-(iv) show the associated magnon spectra. (i) In the symmetric spin-valley-locked phase, both magnon modes are gapped, and the pairing interaction is correspondingly suppressed. (ii) At the spin-canted phase transition both modes become gapless and have identical low-energy dispersion.(iii) In the canted phase where $\theta_0 \neq 0$, the only gapless mode is the Goldstone mode $s-$ associated with the spontaneous symmetry breaking; however, the $a-$ magnon gap is parametrically small in $J/V$ and also contributes appreciably to pairing. (iv) When $\theta_0 \rightarrow \pi/2$ (corresponding to vanishing spin-orbit interaction), the $s-$ mode acquires quadratic dispersion characteristic of an SU(2) ferromagnet. In this limit we show that magnon-mediated interband pairing vanishes by symmetry.}
    \label{fig2}
\end{figure*}

We organize the rest of this article as follows. Section~\ref{sec:spin-canted-phase} describes the phenomenology of the spin-canted phase in rhombohedral graphene, using both an effective Landau-theory description as well as a microscopic model from which we extract the magnon spectrum. Section~\ref{sec:magnon_mediated_interactions} derives the form of magnon-mediated interactions between electrons, and Sec.~\ref{sec:pairing_interaction_main} then investigates the Cooper-pairing problem mediated by those interactions. Finally, Sec.~\ref{sec:Discussion} compares our results to prior experimental observations, makes key testable predictions, and proposes several future directions.

\section{Phenomenology of the spin-canted phase}
\label{sec:spin-canted-phase}

In this section we present the phenomenology of the spin-canted phase that forms the basis of our proposed mechanism for spin-orbit-enabled superconductivity in rhombohedral graphene.  We capture this symmetry-breaking order from both Landau theory and a physically motivated microscopic model; additionally, we will use the latter to study the magnon spectrum. 

\subsection{Landau-theory description}
\label{sec:Landautheory}

Many of the symmetry-breaking metallic phases observed in rhombohedral graphene can be understood within the generalized Stoner ferromagnetism picture, whereby a subset of the four spin and valley flavors are preferentially occupied.  Here we focus on the situation in which symmetry-breaking order can be captured by the spin polarization
$\vec n_\xi$ in valley $\xi = K,K'$. 
 The magnitude $n_\xi$ is essentially the density difference between the majority and minority spin flavors in valley $\xi$, while the direction $\hat{\vec{n}}_\xi$ sets the spin orientations for the majority and minority carriers. Throughout this section we assume that all order parameters are uniform in real-space (i.e., we only consider their $\vec q=0$ components).

A minimal free energy density describing generalized Stoner ferromagnetism in spin-orbit-proximitized devices reads~\cite{zhou2021half} 
\begin{equation}
    {\cal F} = \frac{\kappa_s}{2} \sum_{\xi} \vec n_\xi^2 
    - J \vec n_{K} \cdot \vec n_{K'} + \frac{\lambda_I}{2} \sum_\xi \tau_\xi n^z_\xi + \cdots.
    \label{eq:free_energy_Stoner}
\end{equation}
In the first term---which is invariant under independent SU(2)  spin rotations in the two valleys---$\kappa_s$ encodes a combination of intravalley exchange interactions and kinetic energy.  Near the onset of a Stoner transition driven primarily by these ingredients, $\kappa_s$ turns negative and promotes the development of non-zero order parameters $\vec n_{\xi}$. 
Intervalley exchange processes are modeled by the second term, which breaks the symmetry of the $\kappa_s$ term down to global SU(2) spin rotations. We assume $J>0$, corresponding to ferromagnetic Hund's coupling that favors aligning spins for the two valleys.
In the third term, $\tau_{K,K'} = \pm 1$ is a valley-dependent sign and $\lambda_I$ denotes Ising spin-orbit coupling that favors spins pointing out-of-plane, in opposite directions in the two valleys, denoted as $\pm \hat{\bf z}$.  Ising spin-orbit coupling further reduces the symmetry down to global U(1) spin rotations around the $z$ axis. For concreteness, we assume $\lambda_I \geq 0$ throughout.  Finally, the ellipsis denotes symmetry-allowed terms that are higher-order in $\vec{n}_\xi$ (e.g., cubic and quartic terms). 
Such terms fix the order-parameter magnitudes in the ordered phase and also determine whether the Stoner transition is continuous or first-order~\cite{Coleman2015}. Importantly, our analysis 
applies regardless of the order of the phase transition. 

Let us denote the vectors $\vec{n}_\xi$ by their magnitudes $n_\xi$, their polar angles $\theta_\xi$ measured from the $+\hat{\bf z}$ direction, and their azimuthal angles $\phi_\xi$. The free energy is minimized when the azimuthal angles match, $\phi_+ = \phi_-$, because such configurations optimize the $J$ term without penalizing the $\kappa_s$ or $\lambda_I$ contributions. We further consider solutions where the order-parameter amplitudes agree, 
$n_{K} = n_{K'} \equiv n_0$. Competing orders with unequal $n_{K}$ and $ n_{K'}$~\cite{Arp2024}
can be excluded on phenomenological grounds due to their non-zero valley polarization, which is highly detrimental to zero-momentum pairing.

Imposing the preceding two restrictions yields a free energy density dependent only on $n_0$ and $\theta_{\xi}$:
\begin{equation}
    {\cal F} = \kappa_s n_0^2
    -J n_0^2 \cos (\theta_+ - \theta_-) + \frac{\lambda_I}{2} n_0 \left( \cos \theta_+ - \cos \theta_- \right) + \cdots
    \label{eq:F}
\end{equation}
Next we observe that Eq.~\eqref{eq:F} is minimized---assuming $\lambda_I \neq 0$---by solutions of the form $\theta_+ = \pi - \theta_-$ and $\theta_- \equiv \theta_0$.  Specializing to such configurations yields
\begin{equation}
    {\cal F} = \kappa_s n_0^2
    + J n_0^2 \cos 2 \theta_0 - \lambda_I n_0 \cos \theta_0 + \cdots,
    \label{eq:F2}
\end{equation}
which admits two types of ordered states with $n_0 \neq 0$: $(i)$ For $\lambda_I > 4 J n_0$, the system minimizes its free energy by selecting $\theta_0 = 0$, corresponding to an Ising spin-valley-locked phase that preserves U(1) spin rotation symmetry. $(ii)$ For $\lambda_I < 4 J n_0$, a spin-canted phase emerges with canting angle
\begin{equation}
  \theta_0 = \arccos\left({\frac{\lambda_I }{4J n_0}}\right)
  \label{canting_angle}
\end{equation}
that smoothly increases from 0 towards $\pi/2$ as $\lambda_I$ decreases.
This phase crucially \emph{does} spontaneously break the continuous U(1) spin-rotation symmetry due to the arbitrary orientation of the in-plane moment that develops from canting (see also Fig.~\ref{fig1}), and correspondingly supports a linearly dispersing Goldstone mode. 

For a rough order-of-magnitude estimate, suppose that $\lambda_I \sim 1$ meV and $J \sim 200$ meV nm$^2$~\cite{zhou2021half, dong2023superconductivity,qin2023functional,Shavit2023,Arp2024,Koh2024}. With these parameters our free-energy analysis predicts that spin-canting order sets in when the density difference between majority and minority carriers exceeds $n_0^c \sim 10^{11}$ cm$^{-2}$.  In proximitized rhombohedral bilayers~\cite{Zhang2023, Holleis2025, li2024tunable, Yiran2025} and trilayers~\cite{Patterson2025, LongJu2025}, the normal state hosting the strongest superconductor exhibits polarization densities comparable to $n_0^c$, suggesting spin-canting order as a natural parent phase in agreement with Hartree-Fock simulations~\cite{Koh2024, Koh2024a}.

It is also instructive to consider the SU(2)-invariant limit with $\lambda_I = 0$.  Here the free energy in Eq.~\eqref{eq:free_energy_Stoner} is minimized in the ordered phase for any choice of the polar and azimuthal angles characterizing $\vec{n}_\xi$, provided they are identical for the two valleys.  This phase spontaneously breaks SU(2) symmetry and also exhibits a Goldstone mode, albeit with quadratic rather than linear dispersion.

\subsection{Microscopic model}

For a more microscopic treatment, we now consider the Euclidean action 
\begin{align}
\mathcal{S} =& \sum_{\xi}\int_k\overline{\psi}_{\xi, k} \lp -i\omega -\mu +  \epsilon_{\xi,\vec{k}}  - \frac{\lambda_I}{2}\tau_{\xi}\sigma^z \rp \psi_{\xi, k} 
\nonumber\\
&-\int_q\bigg[\frac{V}{2} \sum_{\xi} \vec s_\xi (q) \cdot \vec s_\xi (-q) 
+J  \vec s_K(q) \cdot \vec s_{K'}(-q)\bigg]. 
\label{S}
\end{align}
In the sums $\xi = K, K'$ is a valley index, while $k = (i\omega,\vec k)$, $q = (i\nu,\vec q)$ denote shorthand Matsubara frequencies and momenta.  Here and below we use Pauli matrices $\sigma^\alpha$ that act in spin space and once again define $\tau_{K,K'} = \pm 1$.  (Spin indices are implicitly summed, though for clarity we explicitly display valley indices.)  Grassman variables $\overline{\psi}_{\xi,k}$ and $\psi_{\xi,k}$ are associated with creation and annihilation operators, respectively, while $\vec s_\xi(q) = \int_{k} \overline{\psi}_{\xi, k+ q} \vec \sigma \psi_{\xi, k}$ is the spin polarization operator for valley $\xi$.  The first line of Eq.~\eqref{S} encodes a chemical potential $\mu$ and spin-independent band energies $\epsilon_{\xi,\vec{k}}$ supplemented by Ising spin-orbit coupling $\lambda_I$.  The second line incorporates phenomenological intravalley ($V>0$) and intervalley ($J>0$) ferromagnetic spin-spin interactions.  In particular, the $V$ term, which is invariant under independent spin rotations in the two valleys, favors polarizing the spins separately in each valley, while $J$ represents a Hund's coupling that favors aligning the spins in the two valleys. As in the free energy discussed in
Sec.~\ref{sec:Landautheory}, spin-canting order arises from the interplay between interaction parameters $V, J$ and the Ising coupling $\lambda_I$.  
 
For convenience, we define the coupling matrix
\be
g_{\xi\xi'} = 
 \lp
 \begin{matrix}
 V & J \\
 J & V
 \end{matrix}
 \rp_{\xi\xi'},
\ee
which allows us to more compactly express the action as 
\begin{align} 
\label{eq:lagrangian2}
\mathcal{S} &= \sum_{\xi}\int_k\overline{\psi}_{\xi, k} \lp -i\omega -\mu + \epsilon_{\xi,\vec{k}} -\frac{\lambda_I}{2}\tau_{\xi}\sigma^z \rp \psi_{\xi, k} 
\nonumber\\
&-\frac{1}{2} \sum_{\xi,\xi'}\int_q g_{\xi\xi'}\vec s_\xi (q) \cdot \vec s_{\xi'}(-q). 
\end{align}
Upon introducing an auxiliary field $\vec{m}_\xi(q)$, we may decouple the interactions via a Hubbard-Stratonovich transformation to arrive at
\begin{align} 
\label{S3}
\mathcal{S} &= \sum_{\xi}\int_k\overline{\psi}_{\xi, k} \lp -i\omega -\mu + \epsilon_{\xi,\vec{k}} - \frac{\lambda_I}{2}\tau_{\xi}\sigma^z \rp \psi_{\xi, k} 
\nonumber\\
&+\int_q\bigg{[}\frac{1}{2}\sum_{\xi,\xi'}g^{-1}_{\xi\xi'}\vec m_\xi (q) \cdot \vec m_{\xi'}(-q) -\sum_{\xi} \vec m_\xi (q) \cdot \vec s_{\xi}(-q)\bigg{]}. 
\end{align}
Next we explore the saddle points of the above action.  

\subsection{Saddle-point analysis}
\label{saddlept}

We specifically seek a uniform saddle-point solution and thus (for now) replace $\vec{m}_\xi(q) \rightarrow \delta(q)(-\frac{\lambda_I}{2} \tau_{\xi}{\bf \hat{z}} + \bar{\vec{m}}_\xi)$.  On the right-hand side we decomposed the $q = 0$ component of $\vec{m}_\xi(q)$ into a piece involving the bare Ising strength and a vector $\bar{\vec{m}}_\xi$ that specifies the net effective field experienced by valley $\xi$.  Motivated by symmetry and the analysis of Sec.~\ref{sec:Landautheory}, we further assume a saddle-point solution of the form
\begin{equation}
    \bar{\vec{m}}_\xi = \bar{m}(\sin\theta {\bf \hat{x}} + \tau_{\xi} \cos\theta{\bf \hat{z}}).
    \label{eq:saddle_point_solution_ansatz}
\end{equation}
Here $\theta$ is the spin-canting angle, with the associated spin profile sketched in Fig.~\ref{fig1}, and we have arbitrarily chosen the in-plane magnetization to orient along the $x$ direction.  Plugging the above ansatz into the action $\mathcal{S}$ and dropping constants independent of $\bar{m}$ and $\theta$ yields
\begin{align}
    \mathcal{S} &\rightarrow \sum_{\xi}\int_k\overline{\psi}_{\xi, k} \lp -i\omega -\mu + \epsilon_{\xi,\vec{k}} - \bar{\vec m}_\xi\cdot \vec \sigma \rp \psi_{\xi, k} 
\nonumber\\
&+\beta A\left(\frac{\bar{m}^2}{V+J} + \frac{2J \bar{m}^2 \cos^2\theta}{V^2-J^2} - \frac{\lambda_I \bar{m} \cos\theta}{V-J}\right) ,
\end{align}
with $\beta$ the inverse temperature and $A$ the system's area.  [The $\beta A$ constant appears through regularizing a factor of $\delta(q = 0)$.]
From the first line it is clear that $\bar{\vec m}_\xi$ indeed captures the total effective field for valley $\xi$ as noted above.  
After integrating out the fermions---which is efficiently done after performing a valley-dependent spin rotation to align the $\sigma^z$ direction with $\bar{\vec{m}}_\xi$ (see also Eq.~\eqref{eq:rotated_basis_def})---we obtain the free energy density
\begin{align}
  \mathcal{F} &= \mathcal{A}(\bar{m}) + \frac{\bar{m}^2}{V+J} + \frac{2J \bar{m}^2 \cos^2\theta}{V^2-J^2} - \frac{\lambda_I \bar{m} \cos\theta}{V-J}.
  \label{eq:free_energy_density_angle}
\end{align}
Here we introduced the function $\mathcal{A}(\bar{m}) = 
\Tr \ln G_{\xi,\sigma}(k)$, where $\Tr$ represents the summation over all isospins, momenta, and frequencies and 
\begin{equation}
  G_{\xi,\sigma}(k) = (i\omega +\mu -\epsilon_{\xi,\vec{k}} + \bar{m}\sigma)^{-1}
  \label{G}
\end{equation} 
is the fermion Green's function in the rotated basis.  Notice that $\mathcal{A}(\bar{m})$ is independent of $\theta$ and, by time-reversal symmetry, only involves even powers of $\bar{m}$.  

Focusing first on the angle-dependent part, the free energy is minimized by nontrivial canting angles 
\begin{equation}
   \theta_0 = \arccos\left[{\frac{\lambda_I (V+J)}{4J \bar{m}}}\right]
   \label{theta0}
\end{equation}
when the argument of the cosine is less than 1; otherwise an uncanted phase with $\theta_0 = 0$ emerges. One can similarly minimize the free energy to determine the optimal amplitude $\bar{m}$.  The result, however, would depend on non-universal band structure details that enter
the function $\mathcal{A}(\bar{m})$.  We adopt an alternative approach and relate $\bar{m}$ to the spin polarization density $n_0$ that can be directly extracted from quantum oscillations data.  In particular, by extremizing the action in Eq.~\eqref{S3} with respect to $\vec{m}_\xi(q)$ and using $|\vec{s}_\xi(q)| = \delta(q) n_0$ as appropriate for our saddle-point solution, we obtain 
\begin{align}
    n_0 = \left|\frac{\bar{m} \sin\theta}{V+J}{\bf\hat{x}} + \frac{\bar{m}\cos\theta-\lambda_I/2}{V-J}{\bf \hat{z}}\right|.
\end{align}
Inserting $\theta_0$ from Eq.~\eqref{theta0} yields the simple relation
\begin{equation} \label{eq: m vs n}
    \bar{m} = n_0(V+J).
\end{equation}
We can, in turn, rewrite $\theta_0$ in terms of $n_0$, recovering 
Eq.~\eqref{canting_angle} derived from Landau theory in Sec.~\ref{sec:Landautheory}. 

\subsection{Magnon fluctuations}
\label{sec:magnons}

The spectrum of magnon fluctuations in the spin-canted phase can be extracted by expanding around the saddle-point solution obtained in the preceding subsection.  With this goal in mind, we now write 
\begin{equation}
    \vec{m}_\xi(q) = \delta(q)\left(-\frac{\lambda_I}{2} \tau_{\xi}{\bf \hat{z}} + \bar{\vec{m}}_\xi\right) + \delta \vec{m}_\xi(q),
\end{equation}
where $\delta \vec{m}_\xi(q)$ are magnon fluctuation fields.  To a good approximation, the soft magnon modes of interest correspond to long-wavelength fluctuations of $\vec{m}_\xi$ that are orthogonal to the saddle-point value of $\bar{\vec{m}}_\xi$.  Technically, fluctuations along $\bar{\vec{m}}_\xi$ and transverse to $\bar{\vec{m}}_\xi$ mix due the low symmetry of the problem, but the former can in principle be safely integrated out, yielding only
quantitative modifications to the low-energy magnon spectrum. For simplicity we 
barbarically project onto the transverse fluctuations\footnote{While this procedure is safe up to leading (linear) order in the fermion-magnon coupling, it throws out corrections to quadratic order in $\delta m$ that become important when considering higher-order magnon exchange processes. In fact, these corrections are crucial in ensuring that the intra-band electron-Goldstone mode coupling vanishes when $\vec q \rightarrow 0$, as expected from general considerations~\cite{Watanabe2014}. \label{footnote_higherorder}} and write
\begin{align}
    \delta \vec{m}_\xi(q) &= \delta m^{\phi}_\xi(q){\bf \hat{y}} + \delta m^{\theta}_\xi(q)(\tau_\xi\cos\theta_0 {\bf \hat{x}}- \sin\theta_0{\bf \hat{z}}).
    \label{deltam_decomp}
\end{align}
Here, $\delta m^\phi_\xi$ parametrizes fluctuations in the in-plane moment orientation about the (arbitrarily chosen) ${\bf \hat{x}}$ direction, 
while $\delta m^\theta_\xi$ captures fluctuations in the canting angle.
Retaining only pieces involving fermions and the magnon fluctuation fields, the action in Eq.~\eqref{S3}
becomes
\begin{align} 
\label{S4}
\mathcal{S} &= \sum_{\xi}\int_k\overline{\psi}_{\xi, k} \lp -i\omega -\mu + \epsilon_{\xi,\vec{k}} - \bar{\vec m}_\xi\cdot \vec \sigma \rp \psi_{\xi, k} 
\nonumber\\
&+\int_q\bigg{\{}\frac{1}{2}\sum_{\xi,\xi'}g^{-1}_{\xi\xi'}\delta \vec m_\xi (q) \cdot \delta \vec m_{\xi'}(-q) 
\nonumber \\
&-\sum_{\xi} \delta \vec m_\xi (q) \cdot [\vec s_{\xi}(-q)-\bar{\vec{s}}_{\xi}]
\end{align}
with $\delta \vec m_\xi (q)$ decomposed as in Eq.~\eqref{deltam_decomp}.
In the last line, $\bar{\vec{s}}_{\xi}= \delta(q)\sum_{\xi'}g^{-1}_{\xi \xi'}(-\frac{\lambda_I}{2}\tau_{\xi'} {\bf \hat{z}} + \bar{\vec{m}}_{\xi'})$ denotes the expectation value of $\vec s_{\xi}(q)$ evaluated at the saddle point obtained earlier; this form makes explicit that the linear-in-$\delta \vec{m}_\xi$ piece couples only to fluctuations in the fermions---which must be the case when expanding around a saddle point.  

Integrating out the fermions to second order in the fermion-magnon interaction yields an effective magnon action
\begin{align} 
\label{Sdeltam}
\mathcal{S}_{\rm mag} &= \frac{1}{2}\int_q\bigg{\{}\sum_{\xi,\xi'}g^{-1}_{\xi\xi'}\delta \vec m_\xi (q) \cdot \delta \vec m_{\xi'}(-q) 
\nonumber \\
&-\sum_\xi \chi_1(q)[\delta m^{\phi}_\xi(q)\delta m^{\phi}_\xi(-q) + \delta m^{\theta}_\xi(q)\delta m^{\theta}_\xi(-q) ]
\nonumber \\
&-2\sum_\xi \chi_2(q) \delta m^\theta_\xi(q) \delta m^\phi_\xi(-q). 
\end{align}
In the first line, for compactness we have not yet explicitly written $\delta \vec{m}_\xi$ in terms of $\delta m_\xi^{\phi,\theta}$, contrary to the remainder of the action.  In the second and third lines we introduced quantities
\begin{align}
  \chi_1(q) &= -\int_p \left[ G_{\xi,+}(p+q)G_{\xi,-}(p) + G_{\xi,-}(p+q)G_{\xi,+}(p) \right], \label{eq:chi_1_definition}
  \\
  \chi_2(q) &= i \int_p \left[ G_{\xi,+}(p+q)G_{\xi,-}(p) - G_{\xi,-}(p+q)G_{\xi,+}(p) \right], \label{eq:chi_2_definition}
\end{align}
dependent on the Green's functions in Eq.~\eqref{G}. 
Appendix~\ref{sec:susceptibility_SI} derives the small-$q$ expressions 
\begin{align}
  \chi_1(q) &\sim \frac{1}{V+J} - \kappa \nu^2 - D {\vec q}^2, ~~~~
  \chi_2(q) \sim - \gamma \nu,
\end{align}
where $\kappa = [4(V+J)\bar{m}^2]^{-1}  $, $D$ is a non-universal constant, and  $\gamma = 
2\kappa \bar{m} $.  
Note that we ignored the self-energy in the electron Green's function, which can shift the dispersion (real part) and blur the Fermi surface (imaginary part). This omission is justified because the self-energy is not expected to alter the qualitative features of the magnons---e.g., their degeneracy, and whether their dispersion is linear or quadratic---that are dictated by general symmetry considerations~\cite{watanabe2020counting}.
With this preparation, we proceed to work out the magnon dispersion. Hund's coupling $J$ embedded in the first line of Eq.~\eqref{Sdeltam} hybridizes fluctuations in the two valleys.  Consequently, it is convenient to pass to a basis of symmetric and antisymmetric valley fluctuations via 
\begin{equation}\label{eq:def_ms_ma}
    \delta m^{\theta,\phi}_K = \frac{\delta m^{\theta,\phi}_s + \delta m^{\theta,\phi}_a}{\sqrt{2}},~~\delta m^{\theta,\phi}_{K'} = \frac{\delta m^{\theta,\phi}_s - \delta m^{\theta,\phi}_a}{\sqrt{2}}.
\end{equation}
Assembling these variables into two-component fields $\delta M_{s,a}(q) = [\delta m^{\phi}_{s,a}(q); \delta m^{\theta}_{s,a}(q)]$ 
enables compactly expressing the magnon action as
\begin{align}
    \mathcal{S}_{\rm mag} &= \frac{1}{2}\int_q\left[\delta M_s^T(-q) \lp
 \begin{matrix}
 Q^\phi_s & \gamma \nu \\
 - \gamma \nu & Q^\theta_s
 \end{matrix}
 \rp\delta M_s(q) + (s \rightarrow a)\right].
 \label{SmagFinal}
\end{align}
Here we have defined 
\begin{align}
&Q^\phi_s = 
\kappa \nu^2 + D {\vec q}^2, \nonumber\\
&Q^\theta_s = 
2\tilde{J}\cos^2\theta_0+\kappa \nu^2 + D {\vec q}^2,\nonumber\\
&Q^\phi_a = 
2\tilde{J} + \kappa \nu^2 + D {\vec q}^2, \nonumber \\
&Q^\theta_a = 
2\tilde{J}\sin^2\theta_0+\kappa \nu^2 + D {\vec q}^2 ,
\label{eq:Q main text}
\end{align}
with $\tilde J = J/(V^2-J^2)$.  Finally, the matrices in Eq.~\eqref{SmagFinal} become diagonal under a rotation of the form $\delta M_{s,a}(q) = U_{s,a}(\nu)[\delta m_{s,a+}(q); \delta m_{s,a-}(q)]$, leading to 
\begin{align}
    \mathcal{S}_{\rm mag} &= \frac{1}{2}\int_q\sum_{b = s,a} \sum_{\eta = \pm} \delta m_{b\eta}(-q) \mathcal{G}^{-1}_{b\eta}(q) \delta m_{b\eta}(q).
 \label{SmagFinal2}
\end{align}
Equation~\eqref{SmagFinal2} describes four branches of magnon modes, labeled $s\pm$ for the valley symmetric sector and $a\pm$ for the valley anti-symmetric sector, with Green's functions 
\begin{align}
    \mathcal{G}_{b\eta}(q) &= 2\left[Q^\phi_b + Q^\theta_b + \eta \sqrt{(Q^\phi_b - Q^\theta_b)^2-(2\gamma\nu)^2}\right]^{-1}.
    \label{Gdef}
\end{align}
The frequency-dependent matrix $U_{s,a}(\nu)$---which we examine further in Appendix~\ref{sec:Umatrix}---determines the weight of the magnon modes on the original $\delta m^{\phi,\theta}_\xi$ fluctuation fields.  

In Appendix~\ref{app:dispersion_of_magnons} we extract the low-lying magnon spectrum from poles of the magnon Green's functions, i.e., frequencies for which $G_{b\eta}^{-1}(q) = 0$. As detailed there, the $a+$ and $s+$ modes exhibit gaps comparable to the Fermi energy $\epsilon_F$. We therefore focus below on the other two modes, $s-$ and $a-$, that are much softer (see also Fig.~\ref{fig2}). For generic canting angles satisfying $0 < \theta_0 < \pi/2$, only the $s-$ mode is gapless.  The structure of the $U_{s}(\nu)$ matrix reveals that, at low frequencies ($\nu \lesssim \frac{\tilde J}{\gamma}\cos^2\theta_0$), $s-$ primarily encodes valley-symmetric $\delta m^\phi$ fluctuations, with subleading weight on $\delta m^\theta$ fluctuations.  Physically, this Goldstone mode reflects long-wavelength azimuthal fluctuations in the orientation of the spontaneously chosen in-plane moment.  The 
 small-${\vec q}$ dispersion reads
\begin{align}
    \omega_{s-}({\vec q}) &= c_s|{\vec q}| + O({\vec q}^2),
    \label{Goldstone}
    \\
    c_s &= \sqrt{\frac{D}{\kappa z_s}} \cos\theta_0 ,
    \label{eq:cs}
\end{align}
where we defined the dimensionless quantity $z_s =  \cos^2\theta_0+\frac{V-J}{2J}$. The linear-dispersion regime is valid for momenta below $q_{\rm max} \sim \frac{1}{V} \sqrt{\frac{J}{D}}\cos \theta_0$ and 
thus shrinks to zero when approaching the SU(2) limit ($\theta_0 \rightarrow \pi/2$), as expected.
Estimating the quantity $VD$ through dimensional analysis as $\sim 1/k_F^2$, one finds $q_{\rm max} \sim \sqrt{\frac{J}{V}} \cos \theta_0 k_F$.  Given that we expect $V/J\sim 10$, the momentum cutoff for linear dispersion is comparable to the Fermi momentum.  See Appendix \ref{app:dispersion_of_magnons} for further numerical investigation of the magnon dispersion.

The dispersion of the $a-$ mode, which at low frequencies primarily involves  valley-antisymmetric $\delta m^\theta$ fluctuations, is given by
\begin{align}
    \omega_{a-}({\vec q}) &= \sqrt{ \Omega_{a-}^2 + c_a^2 {\vec q}^2 },
    \label{eq:a- dispersion}
    \\
    c_a &= \sqrt{\frac{D}{\kappa z_a}} (1+\sin^2\theta_0)^{1/2},
\label{eq:ca}
\end{align}
where we defined  $z_a =(1+\sin^2\theta_0) +\frac{V-J}{2 J }$ and the magnon gap $\Omega_{a-} =  \sqrt{\frac{2\tilde{J}}{ z_a \kappa}}\sin\theta_0$. Up to leading order in $J/V$, we obtain  that $\Omega_{a-} \sim \lp \frac{J\sin\theta_0}{V}\rp \sqrt{\frac{1}{V\kappa }}\sim \lp \frac{J\sin\theta_0}{V}\rp \bar{m}$.  When the polarization densities are comparable to the total electronic density~\cite{Zhang2023, Holleis2025, Patterson2025, li2024tunable, Yiran2025, LongJu2025}, we further have $\bar{m} \sim \epsilon_F$.  Consequently, the $a-$ magnon gap lies well below $\epsilon_F$ in the experimentally relevant regime as claimed.

At $\theta_0 = 0$, the $a-$ mode becomes gapless and degenerate with the $s-$ mode. Note that $\theta_0 = 0$ corresponds to the phase boundary between canted and uncanted order, and hence no symmetries are broken in this limit.  The $s-$ and $a-$ modes are not Goldstone modes, but rather respectively encode critical long-wavelength fluctuations in the in-plane moment along the $y$ and $x$ directions---reflecting proximity to spin-canted order.  

The opposite extreme where $\theta_0 = \pi/2$ corresponds to a ferromagnetic phase in the spin-orbit-free limit $\lambda_I \rightarrow 0$ [recall Eq.~\eqref{canting_angle}]. 
The velocity in
Eq.~\eqref{eq:cs} vanishes here, indicating the onset of quadratic dispersion consistent with a spontaneously broken SU(2) rather than U(1) spin symmetry. For further discussion of this limit see Appendices~\ref{sec:Umatrix} and \ref{app:dispersion_of_magnons}.

\section{Magnon-mediated interactions}
\label{sec:magnon_mediated_interactions}

The magnons analyzed in Sec.~\ref{sec:magnons} mediate electron-electron interactions that can provide a glue 
for superconductivity, as we explore in the following section.  As a preliminary step, here we 
derive the magnon-mediated interactions starting from the action in Eq.~\eqref{S4} that 
features electron-magnon coupling.  It will prove convenient to introduce spin-rotated electron operators via
\be
\psi_{K,k} = e^{-i \frac{\theta_0}{2}\sigma^y}\Psi_{K,k} ~,  ~ \psi_{K',k} = e^{-i \frac{\pi-\theta_0}{2}\sigma^y}\Psi_{K',k}.
\label{eq:rotated_basis_def}
\ee
In the rotated basis, the net field acting on valley $\xi$ electrons becomes 
\begin{equation}
    -\psi_{\xi,k}^\dagger\bar{\vec{m}}_\xi\cdot\vec{\sigma} \psi_{\xi,k} = -\bar{m} \Psi_{\xi,k}^\dagger \sigma^z \Psi_{\xi,k};
\end{equation}
hence for $\Psi_{\xi,k}$, `spin up' (denoted $+$ below) and `spin down' (denoted $-$) respectively label the majority and minority Fermi surfaces for each valley. The electron-magnon coupling in Eq.~\eqref{S4} maps to 
\begin{equation}
    \delta \vec{m}_\xi(q) \cdot \vec{s}_\xi(-q) = \delta m^\phi_\xi(q) S^y_\xi(-q) + \delta m^\theta_\xi(q) S^x_\xi(-q),
\end{equation} 
where $S^{x,y}_\xi(q) = \int_k \overline{\Psi}_{\xi,k+q}\sigma^{x,y} \Psi_{\xi,k}$.

One can formally integrate out magnons from Eq.~\eqref{S4} using the magnon Green's functions constructed in Sec.~\ref{sec:magnons}.  This procedure---which effectively sums an infinite series of diagrams involving the bare electron-magnon coupling---generates four-fermion interactions of the form
\be
\mathcal{S}_{\rm I} = \sum_{\xi_1\xi_2} \int_{kqp} g^{\alpha' \alpha, \beta' \beta}_{\xi_1,\xi_2}(q) \overline{\Psi}_{\alpha'\xi_1, k+q}   \overline{\Psi}_{\beta'\xi_2, p-q }  \Psi_{\beta\xi_2, p}  \Psi_{\alpha\xi_1, k},
\label{SI}
\ee
where $\xi_{1,2}$ are valley indices and $\alpha,\alpha',\beta,\beta'$ are implicitly summed rotated-spin indices that label the majority and minority Fermi surfaces as specified above. 
The scattering amplitudes in $\mathcal{S}_{\rm I}$ satisfy $g^{\alpha' \alpha, \beta' \beta}_{\xi_1,\xi_2}(q) = g^{\beta' \beta, \alpha' \alpha}_{\xi_2,\xi_1}(-q)$ (for trivial reasons) and $g^{\alpha' \alpha, \beta' \beta}_{\xi_1,\xi_2}(q) = [g^{\alpha \alpha', \beta \beta'}_{\xi_1,\xi_2}(-q)]^*$ (reflecting Hermiticity).  They can be classified into two types: intervalley amplitudes with $\xi_1 \neq \xi_2$ and intravalley amplitudes with $\xi_1 = \xi_2$.  Moreover, given the structure of the electron-magnon coupling in the rotated basis, all magnon-mediated interactions invariably transfer electrons between the majority and minority Fermi surfaces for both $\xi_1$ and $\xi_2$.  It follows that the only nontrivial scattering amplitudes are 
\begin{align}
    g^{+-,+-}_{\xi_1,\xi_2}(q)  = [g^{-+,-+}_{\xi_1,\xi_2}(-q)]^* \\
    g^{+-,-+}_{\xi_1\xi_2}(q) = [g^{-+,+-}_{\xi_1,\xi_2}(-q)]^*.
\end{align}
In the remainder of this section we calculate the leading contributions to these scattering amplitudes from the soft magnon modes $s-$ and $a-$. As preparation for the pairing problem analyzed in Sec.~\ref{sec:pairing_interaction_main}, we focus specifically on inter-valley scattering amplitudes where two electrons in the majority Fermi surface are scattered to the minority Fermi surface (see Fig.~\ref{fig1}(c)), or vice versa---e.g., $g^{+-,+-}_{K,K'}(q)$. Appendix~\ref{app:derive_magnon_interaction} details the analysis and considers other scattering amplitudes as well. We treat different canting angle regimes separately, starting with the generic spin-orbit-coupled case where $\theta_0 < \pi/2$.

\subsection{Canting angles $\theta_0 < \pi/2$ \label{sec:canting_new}}

We obtain the magnon-mediated pairing interaction by first integrating out \emph{all} magnons (see Appendix \ref{app:derive_magnon_interaction}), and then isolating the soft-magnon contributions by focusing on the leading frequency and momentum dependence.  We thereby obtain intervalley scattering amplitudes
\begin{align}
\label{eq:g at theta=0}
g^{+-,+-}_{K,K'}(q) & \sim  \frac{1}{2z_s \kappa }\frac{\cos^2 \theta_0}{\nu^2 +c_s^2\vec q^2}  + \frac{1}{2z_a \kappa }\frac{\cos^2 \theta_0}{\nu^2 + c_a^2\vec q^2 + \Omega_a^2}.
\end{align}
The first term on the right-hand side arises from the $s-$ Goldstone mode, which remains gapless throughout the ordered phase, while the second term arises from the low-lying $a-$ mode that becomes gapless at criticality. As a reminder, the $a-$ magnon gap is smaller than $\epsilon_F$ by an $O (J/V)$ factor; see text below Eq.~\eqref{eq:a- dispersion}.
As a result, pair scattering mediated by this mode
features significant retardation, and therefore will be roughly as relevant for pairing as the Goldstone mode.
At the critical point where $\Omega_a \rightarrow 0$, the restored U(1) symmetry enforces a degeneracy of the two gapless modes, which manifests as $c_s = c_a$ and $z_s = z_a$.

Intriguingly, the Goldstone-mediated part of the interaction in Eq.~\eqref{eq:g at theta=0} features a divergence at low frequency and momentum---which may appear counter-intuitive. One might
expect the electron-Goldstone-mode coupling to 
scale linearly with $\vec q$, given that the electron energy can not be perturbed by a $\vec q=0$ Goldstone mode~\cite{Watanabe2014}. Such scaling would in turn generate an extra $\vec q^2$ factor in the numerator of Eq.~\eqref{eq:g at theta=0}, killing the singularity at small $q$, similarly to electron-electron interactions mediated by acoustic phonons which typically take the form $g_{\rm ph}(q) \sim \frac{c_{\rm ph}^2 {\vec q}^2}{\nu^2 + c_{\rm ph}^2 {\vec q}^2}$, with $c_{\rm ph}$ the phonon velocity. The resolution is that the general argument for linear-in-$\vec q$ coupling
only applies to diagonal scattering amplitudes~\cite{Watanabe2014}. In contrast,
our pairing scenario involves off-diagonal (spin-flipping) scattering, and therefore the electron-magnon vertex can be nonvanishing at small $\vec{q}$, consistent with our explicit calculation.
Similar behavior arises for inter-band electron-phonon coupling, which contains one less $\vec q$ factor compared to intra-band processes (see, e.g., Ref.~\onlinecite{borysenko2011electron}). 

Importantly, however, pairing in our scenario does not benefit from the divergent small-$\vec q$ part of the pairing interaction, due to the momentum-space distance between the majority and minority Fermi surfaces.
An intra-band, small-$\vec q$ component can be obtained through virtual excitations from one Fermi surface to another (and back), but such effects also vanish as $\vec q \rightarrow 0$, in line with Adler's theorem showing that long-wavelength Goldstone modes couple weakly to electrons.  This subtle issue will be clarified in a forthcoming work~\cite{followup}.

\subsection{Spin-orbit-coupling-free limit: $\theta_0 = \pi/2$}
\label{sec:magnon_mediated_interactions_FM}

The ferromagnetic phase with vanishing spin-orbit coupling exhibits canting angle $\theta_0 = \pi/2$ and supports a gapless, quadratically dispersing Goldstone mode. Following the procedure outlined above---which is equivalent to simply setting $\theta_0 = \pi/2$ in Eq.~\eqref{eq:g at theta=0}---we obtain
\begin{align}
    g^{+-,+-}_{K, K'}(q) = 0 .
    \label{eq:quadratic_dispersion}
\end{align}
Vanishing of $g^{+-,+-}_{K, K'}$ descends from SU(2) symmetry present in the microscopic Hamiltonian in the $\lambda_I = 0$ limit.  In the ferromagnetic phase considered here, this symmetry is reduced to U(1), corresponding to global spin rotations about the arbitrarily selected spin polarization axis---taken here to be ${\bf \hat{x}}$.  Interactions must therefore preserve the total spin along $x$, which in the rotated basis corresponds to $S^x_{\rm tot} = \sum_\xi \int_k \overline{\Psi}^\dagger_{\xi,k}\sigma^z \Psi_{\xi,k}$. The $g^{+-,+-}_{K,K'}$ scattering processes flip two spins in the rotated basis in clear violation of this conservation law. This symmetry constraint prevents single-magnon exchange from mediating pairing in SU(2) ferromagnets, and provides a promising explanation for spin-orbit-proximitized superconductivity in rhombohedral graphene. 

\section{Pairing from magnons}
\label{sec:pairing_interaction_main}

Can the spin-flipping pair scattering process described above give rise to superconductivity? In this Section, we study this possibility and estimate the corresponding superconducting $T_c$. 

Due to the spin-flipping nature of the magnon-mediated interaction in
Eqs.~\eqref{eq:g at theta=0}, the pairing process couples the superconducting gaps in the two (majority and minority) spin flavors, which we denote by $\Delta_+$ and $\Delta_-$, respectively. The linearized gap equation is given by
 \begin{align}
 \label{eq:self-consistency simplified}
& \Delta_+(k) = - \sum_{k'} g^{-+,-+}_{K,K'} (k-k')\frac{\Delta_-(k')}{\omega'^2 +\epsilon_{\vec k',-}^2},\\
& \Delta_-(k) = - \sum_{k'}g^{+-,+-}_{K,K'} (k-k')\frac{\Delta_+(k')}{\omega'^2 +\epsilon_{\vec k',+}^2},
 \end{align}
 where $k = (\omega,\vec k)$. 
 Below we focus on the momentum-independent, $s$-wave pairing channel (see however footnote \ref{footnote_swave}), justified by the smooth momentum dependence of the pairing interaction. Due to the interband nature of the pairing, the coupled equations \eqref{eq:self-consistency simplified} admit a momentum-independent solution regardless of the sign of $g^{-+,-+}_{K, K'}$. In our case, $g^{-+,-+}_{K, K'} > 0$ represents a repulsive interaction, and pairing can thus occur in a channel with opposite signs of $\Delta_+$ and $\Delta_-$, similar to $s_{+-}$ pairing in iron-based superconductors (see, e.g., Ref.~\onlinecite{mazin2008unconventional}).
 
To proceed analytically, we apply the trick conventionally used in solving the BCS problem: replacing the full pairing interaction with a frequency-independent version that exhibits a hard cut-off at the characteristic magnon energy $\Omega_0$ (analogous to the Debye energy in conventional phonon-mediated pairing). 
\addELH{We first focus on the critical point, $\theta_0=0$, where pairing is the strongest. Here the two magnon velocities are identical by symmetry, $c_s = c_a = c$, with $c \approx \sqrt{\frac{2 J D}{V\kappa}}$
in the limit $V \gg J$.} The characteristic magnon energy is estimated as $\Omega_0 = c k_0$, where $k_0$ is a characteristic scale for the momentum transfer $\vec k-\vec k'$ in the spin-flipping scattering process. This scale is upper-bounded by $2 k_F^+$, with $k_F^+$ the Fermi momentum for the majority Fermi surface. However, in rhombohedral graphene the momentum-space separation between majority and minority pockets can be several times smaller than $k_F^+$, due in particular to trigonal warping (e.g., see Ref.~\onlinecite{Koh2024}).

With these considerations in mind, we apply the following replacement:
\begin{align}
 \frac{1}{(\omega-\omega')^2 + c^2 (\vec k-\vec k')^2} &\rightarrow \frac{\Theta (\Omega_0-|\omega'|)\Theta (\Omega_0-|\omega|)}{\Omega_0^2} ,\nonumber\\ \Delta_\pm(\omega,\vec k) &\rightarrow  \Delta_\pm^{(0)} \Theta (\Omega_0-|\omega|),\label{eq:Omega0}
\end{align}
which reduces the linearized gap equation to an analytically soluble form,
 \begin{align}\label{eq:gap equation soluable}
	& \Delta_+^{(0)} = -g_{\rm eff}\nu_- \Delta_-^{(0)}  \log \frac{\Omega_0}{T_c}, \\
	& \Delta_-^{(0)} = -g_{\rm eff} \nu_+ \Delta_+^{(0)}  \log \frac{\Omega_0}{T_c}.
    \label{eq:gap equation soluable_2}
\end{align}
\approve{Here, $g_{\rm eff}$ is the pair scattering amplitude evaluated at the smallest momentum connecting the two Fermi surfaces: $g_{\rm eff}=g_{K,K'}^{-+,-+}(k_0)$ with $k_0=k_{F+} -k_{F-}$.}

\approve{At the canting critical point $\theta_0=0$, the $a-$ mode is gapless and degenerate with the $s-$ mode. Both modes contribute constructively and the pairing interaction is maximized, with
\be\label{eq:g_eff theta=0}
g_{\rm eff} \approx \frac{4 J}{V \kappa \Omega_0^2} \approx \frac{2}{D k_0^2}.
\ee 
Equation~\eqref{eq:g_eff theta=0} holds when the magnons used in the pair-scattering process 
disperse linearly. 
This assumption holds 
below a 
momentum cutoff 
$q_{\rm max}\sim O(\sqrt{J/V} ) k_F$.   
(At larger $|{\vec q}|$, magnons disperse quadratically; Appendix~\ref{app:derive_magnon_interaction} provides 
more general expressions for scattering amplitudes valid also in that regime.) Our assumption that pair scattering transfers a momentum $|{\vec q}| \lesssim q_{\rm max}$ is reasonable because
(1) $V/J$ is expected to be $\sim O(10)$ in rhombohedral graphene devices (see Appendix \ref{app:dispersion_of_magnons}), so that $\sqrt{V/J}\sim3$ is not too large; and (2) due to trigonal warping, pair scattering between minority and majority pockets can occur with momenta a few times smaller than $k_F$.}

\addELH{Defining an effective density of state $\nu_{\rm eff } = \sqrt{\nu_+\nu_-}$, the superconducting $T_c$ solved from Eqs.~\eqref{eq:gap equation soluable} and \eqref{eq:gap equation soluable_2} reads
\be\label{eq:Tc}
 T_c = \Omega_0 \exp{\lp -\frac{1}{g_{\rm eff} \nu_{\rm eff}} \rp  }.
\ee
This result enables a crude $T_c$ estimate \addELH{near the critical point}. 
The magnon energy $\Omega_0$ is proportional to a small dimensionless ratio $\sqrt{J/V}$ times the Fermi energy (again assuming that $\bar{m} \sim \epsilon_F)$.
Appendix \ref{app:dispersion_of_magnons} provides an experimentally motivated discussion that estimates $\Omega_0 \sim 1$ meV.} The dimensionless coupling constant $g_{\rm eff} \nu_{\rm eff}$
is independent of the ratio $J/V$ and only involves band structure quantities---it is expected to be of $O(1)$ from dimensional analysis, but will depend on the fermiology and density of states of a specific system. The superconducting $T_c$ predicted in our scenario can therefore reach a substantial fraction of $\Omega_0$. Given that we neglected the competing effect of repulsive Coulomb interactions, this estimate should be understood as an \emph{upper bound} on $T_c$. Indeed, in lightly doped rhombohedral bilayers and trilayers the reported critical temperatures~\cite{Zhang2023, Yiran2025, Holleis2025, Patterson2025, li2024tunable, LongJu2025} for the spin-orbit enabled superconductors are only in the range $k_B T_c \sim 10-50 ~ \mu$eV.

\addELH{Moving away from the critical point, pairing is suppressed by 
the $\cos^2 \theta_0$ numerator in Eq.~\eqref{eq:g at theta=0}
and the opening of a gap in the anti-symmetric mode. 
For a generic $\theta_0$, we have
\be\label{eq:g_eff}
g_{\rm eff} \approx \frac{2 J \cos^2\theta_0}{V \kappa} \left( \frac{1}{c_s^2 k_0^2} + \frac{1}{\Omega_{a-}^2 + c_a^2 k_0^2} \right) .
\ee 
The first and second terms respectively correspond to the contributions of the $s-$ and $a-$ modes. }
\addELH{Here the cutoff frequency scale $\Omega_0$ should be determined separately for the two magnon modes: $\Omega_0 \sim c_sk_0$ for the $s-$ mode, and $\Omega_0 \sim \sqrt{\Omega_{a-}^2+c_a^2k_0^2}$ for the $a-$ mode. However, when $\theta$ is not too close to $\pi/2$ these two frequencies are comparable and a few times below the Fermi energy\footnote{Our numerics in Appendix \ref{app:dispersion_of_magnons} check this claim and show that the magnon energy is of order $\sim  1$meV at $q_{\rm{max}}$, indeed a few times smaller than $E_F$.}, such that the pairing interaction exhibits some retardation. At $\theta_0 = \pi/2$, the pairing interaction 
vanishes due to the symmetry restrictions higlighted in Sec.~\ref{sec:magnon_mediated_interactions_FM}.}

\section{Discussion}
\label{sec:Discussion}

Our theory admits several experimentally testable predictions. Perhaps the most significant and unique one is the dependence of superconductivity on Ising spin-orbit coupling strength $\lambda_I$. Qualitatively, 
the pairing interaction in our theory vanishes by symmetry when $\lambda_I$ is switched off, which can elegantly explain the observed superconducting phases in TMD-proximitized rhombohedral bilayer/trilayer systems that rely on induced spin-orbit coupling~\cite{Zhang2023, Holleis2025, Yiran2025, li2024tunable, Patterson2025,LongJu2025}. Our theory further naturally accommodates two seemingly contradictory dependences on spin-orbit coupling, which cannot be easily understood in a conventional (e.g., phonon-mediated) pairing mechanism:
\begin{enumerate}
\item {\bf The phase diagram area populated by superconductivity shrinks upon increasing spin-orbit coupling.} In our scenario the pairing glue---soft magnons---relies on spin-canting order in the normal state.  As $\lambda_I$ increases, however, canting order eventually loses in favor of an Ising spin-valley-locked phase; recall Sec.~\ref{sec:Landautheory}.  It is thus natural to anticipate that the spin-canted phase occupies a progressively smaller region of the phase diagram as $\lambda_I$ 
rises, as seen in Hartree-Fock studies~\cite{Koh2024, Koh2024a}.  Our theory then predicts that superconductivity should also occupy a diminished area. 

\item { \bf The critical temperature $T_c$ is enhanced upon increasing the spin-orbit coupling.}  This prediction follows directly from our analysis of the effective pairing strength in
Eq.~\eqref{eq:g_eff}, which contains a $\cos^2\theta_0$ prefactor that is proportional to $\lambda_I^2$ through Eq.~\eqref{canting_angle}.
It is then natural that turning on $\lambda_I$ increases the optimal 
$T_c$ emerging from a spin-canted normal state, even though the latter's phase space diminishes.
\end{enumerate}
Remarkably, Ref.~\onlinecite{Yiran2025} observed both of these
trends in bilayer devices, where the strength of Ising spin-orbit interaction was controlled through engineering the twist angle at the WSe$_2$-graphene interface~\cite{Li2019, David2019, Naimer2021}. 

Another nontrivial prediction concerns the dependence of superconductivity on Fermi surface structure. We showed that, due to its interband nature, our proposed pairing mechanism crucially relies on the presence of minority Fermi surfaces. 
More precisely, the dimensionless coupling constant in Eq.~\eqref{eq:Tc} depends on the minority carriers through (the square root of) their density of states. If modeling the minority pockets using a parabolic dispersion,
we therefore expect superconductivity 
to be relatively insensitive to a decrease of minority carrier density, until it gets
abruptly suppressed upon vacating the minority pockets.
This prediction matches the generic behavior observed in bilayer and trilayer 
graphene, where prominent superconducting phases only appear in regions hosting minority pockets~\cite{Zhang2023,Holleis2025,Yiran2025,li2024tunable,Patterson2025,LongJu2025}. Even more strikingly, in trilayers superconductivity abruptly terminates near Lifshitz transitions marking the disappearance of minority carriers~\cite{Patterson2025, LongJu2025}.

Direct detection of the spin-canting order central to our theory (e.g., through SQUID measurements, or by tracking the behavior of phase boundaries with applied magnetic fields in compressibility data) poses arguably the simplest falsifiable check for our proposed explanation. Perhaps the most striking consequence of pairing from a spin-canted normal state, due to its in-plane magnetic moment, is that the extent of superconductivity near phase boundaries to a non-magnetized competing phase can \emph{increase}\footnote{This behavior is not universal because $B_\parallel$ might also have competing pair-breaking effects, in particular as it breaks U(1) spin rotation symmetry and opens up a gap in the magnon dispersion, thus reducing the pairing interaction. Nevertheless, observing such an increase in phase space would constitute evidence for an in-plane magnetic moment in the superconducting phase.}
under the application of a small in-plane field. This behavior was recently reported in proximitized rhombohedral trilayers~\cite{Patterson2025}, along one of the phase boundaries that delimit the strongest spin-orbit-enabled superconductor.

Our pairing scenario also has ramifications for depairing by in-plane magnetic fields $B_{\parallel}$. Conventional in-plane-field depairing mechanisms include orbital effects and warping of the Fermi surfaces due to an interplay with Rashba spin-orbit coupling~\cite{Gorkov2001} (which we have neglected so far but discuss further below).  
Since the pairing glue in our scenario is based on magnons, one might suspect a non-trivial magnetic-field dependence of $T_c$. Indeed, $B_\parallel$ significantly affects the magnon spectrum near $\vec q = 0$ as
it explicitly breaks the U(1) spin-rotation symmetry that the spin-canting phase breaks spontaneously---\addELH{thereby generating a gap $\Delta_{\rm mag} \sim \sqrt{|B_\parallel|}$ for the Goldstone magnon mode (see Appendix \ref{app:B-dependent_magnon_gap} for a derivation).
Interestingly, this $\sqrt{B_\parallel}$ magnon gap results in \addELH{an unconventional behavior of the critical temperature reduction, which scales as $|B_\parallel|$ at weak fields instead of the traditional Ginzburg-Landau expectation $T_c - T_c^{(0)} \sim B_\parallel^2$. At the canting transition, the gap scaling is modified to $\Delta_{\rm mag} \sim |B_\parallel|^{1/3}$---resulting in an unusual power-law for depairing, with $T_c - T_c^{(0)} \sim B_\parallel^{2/3}$. Magnon-induced depairing is therefore most effective at the canting transition (where pairing is strongest) because of the enhanced susceptibility towards opening a magnon gap at weak fields. These features constitute unique experimental predictions of our pairing scenario. We note that} while our mechanism implies a greater depairing effect of in-plane magnetic fields compared to, say, a phonon-mediated scenario,
its observability in a given material platform will depend on whether orbital depairing, Rashba or magnon effects are predominant. Nevertheless, the dependence of the magnon gap on magnetic field and spin-orbit coupling could be imaged experimentally using, e.g., microwave resonator techniques~\cite{Zhang2014magnons, Tabuchi2014magnons, Boettcher2024}.} 

As alluded to above, Rashba spin-orbit coupling $\lambda_R$ generically accompanies Ising spin-orbit interaction in graphene multilayers proximitized by a TMD.  And like in-plane magnetic fields, $\lambda_R$ explicitly breaks U(1) spin symmetry, gapping the Goldstone magnon even at $B_{\parallel} = 0$.  The bare Rashba energy scale is believed to be relatively weak, $\lambda_R \sim 1 ~ \rm{meV}$, in graphene devices proximitized by WSe$_2$ or WS$_2$~\cite{Gmitra2016, Wang2016, Gmitra2017, Khoo2017, Wang2019, Island2019, Amann2022, Li2019, David2019, Naimer2021}.  Crucially, however, 
the effective Rashba spin splitting in the low-energy bands can be strongly suppressed by the applied displacement field and 
the fermiology of the pockets relevant for pairing. In principle, measuring devices with variable $\lambda_R$, and ideally other parameters fixed, would provide another falsifiable check for our theory.

\addELH{The soft magnons mediating Cooper pairing in this work are subject to Landau damping due to magnon decay into electron-hole pairs near the Fermi surface. A legitimate concern is whether such Landau damping---ignored so far in our analysis---will wash out the magnon-mediated pairing interaction. In Appendix~\ref{sec:landau_damping} we find that Landau damping in our model modifies the spin susceptibility $\chi_1 $ by a term linear in $|\nu|$, thereby suppressing the pairing interaction.   However, since its impact is $O(|\nu|)$ and vanishes at $\nu=0$, there always exists a frequency scale---which we estimate through dimensional analysis 
to be of $O(\frac{J}{V} E_F)$---below which the impact of damping is subleading. Given that the pairing mechanism mainly uses magnons with frequency below $O(\frac{J}{V} E_F)$, we conclude that the impact of damping on the pairing interaction will be limited and that our analysis of pairing remains valid.}

\addELH{Throughout our work we also neglected the (material-dependent) form factors arising from projecting the pairing interaction into the relevant electronic bands. While these form factors carry some momentum dependence, our (projected) intervalley pairing interaction
will remain of the same sign along the 
Fermi surface due to orbital time-reversal symmetry. We therefore do not expect the pairing symmetry to be altered: generally, introducing nodes in the gap function costs condensation energy, and is only preferred if the pairing interaction has a sign alternation along the Fermi surface.}

\addELH{We emphasize that our scenario
aims to explain the appearance of spin-orbit-enabled superconductivity in TMD-proximitized devices, where the normal state is valley balanced. We therefore do not claim any relation to the intriguing superconducting phase found in quarter-metal regions 
of the penta-, tetra- and hexa-layer graphene phase diagrams~\cite{han2025signatures, morissette2025superconductivityanomaloushalleffect}, where spin and valley are assumed to be both polarized.
Neither do we claim a connection to 
the superconductivity found in pristine bilayer and trilayer graphene (without a TMD layer). Our perspective, rooted in recent measurements~\cite{Yiran2025,LongJu2025,Patterson2025}, is that superconductivity in graphene with and without a TMD substrate likely 
arises from different mechanisms.
Indeed, superconductivity in TMD-proximitized devices emerges in regions of the phase diagram where it is absent in ``bare” devices---whereas the superconducting phases present in the bare devices seem to be suppressed (or completely absent) with a TMD substrate.}

We conclude with some speculation on extensions of our theory.  Analyzing the favorability of spin-canting order in systems composed of more than three graphene layers, e.g., from a Hartree-Fock angle, would provide valuable insight into the wider applicability of our proposed pairing mechanism. It is also interesting to explore pairing mediated by Goldstone modes associated with other broken-symmetry orders~\cite{Kozii2022}, most notably intervalley coherence that also naturally emerges in graphene multilayers~\cite{Chatterjee2022, You2022, Ming2023, Huang2023, Koh2024,  Zhumagulov2023, Wang2024, Koh2024a, Das2024, Zhumagulov2024, Vituri2025, liu2024visualizingincommensurateintervalleycoherent, liao2024promotingimagingintervalleycoherent}. A particularly interesting possibility---which combines the two different types of continuous symmetry breaking relevant for graphene systems---consists of intervalley coherent orders that also exhibit spin canting~\cite{Koh2024a}. Such composite order can lead to generalized quarter-metal states that comprise a single, non-degenerate majority Fermi surface; similarly to the ideas developed in this work, our magnon-mediated pairing mechanism could potentially mediate Cooper pairing in this setting. Interestingly, one of the new superconducting regions in proximitized Bernal bilayer graphene~\cite{Zhang2023} exhibits fermiology consistent with this description.

Another promising direction consists of investigating the spin-canting critical point ($\theta_0 \rightarrow 0$) more closely. There, the two gapless ($a-$ and $s-$) modes may mediate pairing interactions with an interesting frequency dependence beyond BCS theory, which will be investigated in upcoming work~\cite{followup}. This endeavor is further motivated by a recent experiment suggesting that superconductivity in spin-orbit-proximitized trilayers occurs on top of a phase transition where spin-canting order sets in~\cite{Patterson2025}.

Looking beyond graphene, can one identify a broader set of candidate materials for which Cooper pairing emerges from magnon exchange processes boosted by spin-orbit interaction?  Ferromagnetic bulk materials composed of weakly coupled layers may be fruitful in this regard. In particular, the magnetic order could inherit a spin-canted structure upon introducing spin-orbit coupling. For systems realizing more complex
 magnetic order (e.g., antiferromagnetism), can magnon exchange still provide a fruitful pairing mechanism---possibly leading to superconducting order parameters beyond a pseudospin-singlet $s$-wave?  Finally, can one uncover excitonic analogues of our magnon-mediated pairing scenario? 

\section*{Acknowledgments}

We are grateful to Trevor Arp, Erez Berg, Nick Bultinck, Andrey Chubukov, Julian Ingham, Long Ju, Hyunjin Kim, Patrick Lee, Leonid Levitov, Cyprian Lewandowski, Stevan Nadj-Perge, Caitlin Patterson, Gal Shavit, Owen Sheekey, Tomohiro Soejima, Alex Thomson, Yaar Vituri,  Andrea Young, and Yiran Zhang for insightful discussions.
We want to thank in particular Alex Thomson and Johnson Guanyao Chen for pointing out a mistake in the previous version of this manuscript, where an intraband two-magnon pairing interaction diverging as $q \rightarrow 0$ was obtained. This contribution is in fact canceled by higher-order terms in the magnon-fermion coupling; see footnote \ref{footnote_higherorder}.
Z.~D.~and \'E.~L.-H.~are supported by the Gordon and Betty Moore Foundation’s EPiQS Initiative, Grant GBMF8682. Portions of this work were supported by the U.S.~Department of Energy, Office of Science, National Quantum Information Science Research Centers, Quantum Science Center (derivation of magnon spectrum and magnon-mediated interactions, J.~A.). 
Additional support was provided by the Caltech Institute for Quantum Information and Matter, an NSF Physics Frontiers Center (NSF Grant PHY-2317110).

\bibliography{biblio}
	
\appendix

\begin{widetext}
 
\section{Calculation of spin susceptibilities}\label{sec:susceptibility_SI}

Below we calculate the transverse spin susceptibilities $\chi_{1}$ and $\chi_{2}$ defined in Eqs.~\eqref{eq:chi_1_definition} and~\eqref{eq:chi_2_definition}:
\begin{align}
\chi_{1}(i\nu,\vec q)
=& -\int_{p} \left[ G_{\xi,+}(p+ q)G_{\xi,-}( p) + G_{\xi,-}( p+ q)G_{\xi,+}( p) \right] \nonumber\\
=&  -\int_{\vec p} T\sum_{\omega} \frac{1}{i\lp \omega +\nu\rp -\epsilon_{\xi,\vec p+\vec q}+\bar{m}}\frac{1}{i\omega -\epsilon_{\xi,\vec p}-\bar{m}} + (\nu\rightarrow -\nu, \vec q \rightarrow -\vec q), \nonumber\\
=&  \int_{\vec p} \frac{f(\epsilon_{\xi,\vec p+\vec q} -\bar{m})-f(\epsilon_{\xi,\vec p}+\bar{m})}{i\nu -\lp\epsilon_{\xi,\vec p+\vec q}-\epsilon_{\xi,\vec p}\rp +2\bar{m}}+ (\nu\rightarrow -\nu, \vec q \rightarrow -\vec q),\nonumber\\
=&~ C - \kappa \nu^2 - D \vec{q}^2 , \label{eq:chi_1_result}\\
\chi_{2}(i\nu,\vec q)
=& i \int_{p}[G_{\xi,+}( p+ q ) G_{\xi,-}( p) - G_{\xi,-}( p+ q) G_{\xi,+}(p)]\nonumber \\
=&  i\int_{\vec p} T\sum_{\omega} \frac{1}{i\lp \omega +\nu\rp -\epsilon_{\xi,\vec p+\vec q}+\bar{m}}\frac{1}{i\omega -\epsilon_{\xi,\vec p}-\bar{m}} - (\nu\rightarrow -\nu, \vec q \rightarrow -\vec q), \nonumber\\
=&  -i \int_{\vec p} \frac{f(\epsilon_{\xi,\vec p+\vec q} -\bar{m}) - f(\epsilon_{\xi,\vec p}+\bar{m})}{i\nu -\lp\epsilon_{\xi,\vec p+\vec q}-\epsilon_{\xi,\vec p}\rp +2\bar{m}} - (\nu\rightarrow -\nu, \vec q \rightarrow -\vec q),\nonumber\\
=&~ - \gamma \nu + O(|\vec{q}|^3, \nu^3). \label{eq:chi_2_result}
\end{align}
In the last line of Eqs.~\eqref{eq:chi_1_result} and \eqref{eq:chi_2_result} we have performed an expansion to leading order in frequency and momentum \addELH{(i.e., when the corresponding terms are small compared to $\bar{m})$}. While $\chi_1(i\nu, \vec q)$ is an even function of its arguments, $\chi_2(i\nu, \vec q)$ is an odd function. 
We further expect the absence of any linear-in-$\vec{q}$ term in $\chi_2$ due to the model's rotation symmetry, together with analyticity of  $\chi_2(q)$; the latter property precludes a $\sim | \vec q|$ contribution. \addELH{We note that $\chi_1$ will inherit a non-analytical term $\sim |\nu|$ for $|\vec q| > k_F^+-k_F^-$ due to Landau damping, as detailed in Appendix~\ref{sec:landau_damping}.}

The expression for the parameter $D$ is complicated and depends on the detail of band dispersion, so we do not write it down explicitly. On the other hand, the parameters $C$, $\kappa$ and $\gamma$ can be computed straightforwardly from the $\vec q = 0$ limit of the above expressions:
\begin{align}
    \chi_1(i \nu, 0) &= \int_{\vec p} \frac{f(\epsilon_{\xi,\vec p} -\bar{m})-f(\epsilon_{\xi,\vec p}+\bar{m})}{i \nu + 2 \bar{m}} + (\nu \rightarrow - \nu) \nonumber \\
    &\approx \int_{\vec p} \frac{f(\epsilon_{\xi,\vec p} -\bar{m})-f(\epsilon_{\xi,\vec p}+\bar{m})}{2 \bar{m}} \left \{ \left( 1 - \frac{i \nu}{\bar{2 m}} - \frac{\nu^2}{4 \bar{m}^2} + \ldots \right) +  \left( 1 + \frac{i \nu}{2\bar{m}} - \frac{\nu^2}{4 \bar{m}^2} + \ldots \right) \right\} \nonumber \\
    &= \frac{n_0}{\bar{m}} - \frac{n_0}{4\bar{m}^3} \nu^2,
\end{align}
where $n_0 = \int_{\vec p} \left[ f(\epsilon_{\xi,\vec p} -\bar{m}) -f(\epsilon_{\xi,\vec p}+\bar{m}) \right]$ denotes the polarization density, i.e., the density difference between the majority and minority spin flavors in each valley. Using the same low-frequency expansion we obtain
\begin{align}
    \chi_2(i \nu, 0) &= -i \int_{\vec p} \frac{f(\epsilon_{\xi,\vec p} -\bar{m})-f(\epsilon_{\xi,\vec p}+\bar{m})}{i \nu + 2 \bar{m}} - (\nu \rightarrow - \nu) \approx - \frac{n_0}{2 \bar{m}^2} \nu. 
\end{align}
From these results we extract coefficients
\be \label{eq:parameters}
C =\chi_1(0,0) = \frac{n_0}{\bar{m}}=\frac{1}{V+J},\quad 
\kappa 
= \frac{n_0}{4\bar{m}^3} = \frac{1}{4 (V+J) \bar{m}^2}, \quad \gamma = \frac{n_0}{2\bar{m}^2} = \frac{1}{2 (V+J) \bar{m}} = 2\kappa \bar{m} .
\ee
Above we used the saddle-point condition from Eq.~\eqref{eq: m vs n} to simplify $C$, $\kappa$ and $\gamma$, leading to the expressions quoted in the main text.

\section{Diagonalization of magnon action}
\label{sec:Umatrix}

The magnon action in Eq.~\eqref{SmagFinal} takes the diagonal form in Eq.~\eqref{SmagFinal2} under a basis rotation 
\begin{align}
    \delta M_{s,a}(q) =  \left[\begin{matrix}
 \delta m^{\phi}_{s,a}(q)  \\
 \delta m^{\theta}_{s,a}(q) 
 \end{matrix} \right] = U_{s,a}(\nu) \left[\begin{matrix}
 \delta m_{s,a+}(q)  \\
 \delta m_{s,a-}(q) 
 \end{matrix} \right].
\end{align}
Our aim is to explore low-frequency properties of the matrix $U_{s,a}(\nu)$, which is composed of (right) eigenvectors of the $2\times2$ matrices appearing in Eq.~\eqref{SmagFinal}.  For generic canting angles $\theta_0$ not too close to $0$ or $\pi/2$, the only low-energy magnon mode corresponds to the $s-$ branch.  The corresponding eigenvector is given by
\begin{equation}
    \Phi_{s-}(\nu) = \frac{1}{\sqrt{\mathcal{N}}}\left[\begin{matrix}
   \tilde J \cos^2\theta_0 + \sqrt{(\tilde J \cos^2\theta_0)^2 - (\gamma \nu)^2}\\
 \gamma \nu 
 \end{matrix} \right]
\end{equation}
with $\mathcal{N}$ a normalization factor.
For low frequencies satisfying $\gamma \nu \lesssim \tilde J \cos^2\theta_0$, to a good approximation only the upper entry survives, and one can simply take $\Phi_{s-}(\nu)  \approx \left[\begin{matrix}
   1\\
 0
 \end{matrix} \right]$. That is, at low frequencies the Goldstone mode has weight primarily on the $\delta m_s^\phi(q)$ fluctuation field.  

For $\theta_0 = 0$, the $a-$ branch joins $s-$ as a second gapless mode. The $a-$ eigenvector is
\begin{equation}
    \Phi_{a-}(\nu) = \frac{1}{\sqrt{\mathcal{N}'}}\left[\begin{matrix}
   \tilde J - \sqrt{\tilde J^2 - (\gamma \nu)^2}\\
 -\gamma \nu 
 \end{matrix} \right],
\end{equation}
where $\mathcal{N}'$ again represents a normalization factor.  At low frequencies satisfying $\gamma \nu \lesssim \tilde J^2$, the upper entry is now subdominant, and hence in this regime we have $\Phi_{a-}(\nu)  \approx \left[\begin{matrix}
   0\\
 1
 \end{matrix} \right]$.  The gapless $a-$ mode thus has weight primarily on the $\delta m_a^\theta(q)$ fluctuation field.  

Finally, at $\theta_0 = \pi/2$, one must consider both the $s-$ and $s+$ branches, with eigenvectors
\begin{equation}
    \Phi_{s-}(\nu) = \frac{1}{\sqrt{2}}\left[\begin{matrix}
    1\\
 - i{\rm sgn}(\nu)
 \end{matrix} \right],~~~~~~ \Phi_{s+}(\nu) = \frac{1}{\sqrt{2}}\left[\begin{matrix}
    1\\
 +i{\rm sgn}(\nu) 
 \end{matrix} \right].
 \label{PhiEigenvectors}
\end{equation}
Contrary to the cases examined above, these modes admit equal weight on the $\delta m^\phi_{s}$ and $\delta m^\theta_{s}$ fields (for any nonzero frequency).  We stress that one should not view $s+$ as a mode that gradually evolves from gapped to gapless as $\theta_0$ approaches $\pi/2$.  Rather, the $s+$ abruptly acquires a low-frequency, low-momentum pole precisely at $\theta_0 = \pi/2$.  See the following appendix for more discussion. 

\section{Dispersion of magnons}
\label{app:dispersion_of_magnons}

In the main text, we obtained magnon propagators
 \begin{align}\label{eq:g_eigenvalue_appendix}
    \mathcal{G}_{b\eta}(q) &= 2\left[Q^\phi_b + Q^\theta_b + \eta \sqrt{(Q^\phi_b - Q^\theta_b)^2-(2\gamma\nu)^2}\right]^{-1}
\end{align}
with $b=s,a$ and $\eta = \pm$. We now solve for the low-energy magnon spectrum by examining the poles of these propagators. Specifically, we focus on the $s-$ and $a-$ branches since the former is gapless throughout the spin canted phase while the latter becomes gapless at the spin-canted phase transition.  We generally ignore the gapped $s+$ and $a+$ branches (but see below for the special case of $\theta_0 = \pi/2$).  Solving the poles at small $q$ for the $s-$ mode, we find the following dispersion relation:
  \begin{align}\label{eq:magnon-dispersion_sup}
 \omega_{s-} &=
 \begin{cases}
 c_s |\vec q| +O(|\vec q|^2),
 \quad &0 \leq \theta_0< \frac{\pi}{2}
\\
\frac{D}{\gamma}|\vec q|^2+O(|\vec q|^4),
 \quad &\theta_0= \frac{\pi}{2}
 \end{cases}
 .
 \end{align}
Away from $\theta_0 = \pi/2$, this mode disperses linearly with magnon velocity
\begin{equation}
    c_s = \lp \frac{D}{\kappa  z_s} \rp^{1/2} \cos\theta_0 , \quad z_s =  \cos^2\theta_0+\frac{V-J}{2 J }.
\end{equation}
Similarly, for the $a-$ mode we obtain
\begin{equation}
   \omega_{a-} =  
\sqrt{\Omega_{a-}^2 +c_a^2 |\vec q|^2} + O(|\vec q|^4).
\end{equation}
The associated velocity is
\begin{align}
c_a &= \lp \frac{D}{\kappa z_a} \rp^{1/2} \lp 1+\sin^2\theta_0\rp^{1/2}, \quad z_a = (1+\sin^2\theta_0) +\frac{V-J}{2 J }; \label{eq:c and Z}
\end{align}
moreover, the $\vec{q} = 0$ magnon gap reads
\begin{align}
\Omega_{a-}= \lb \frac{\tilde{J}}{\kappa}(1+\sin^2\theta_0) +\frac{\gamma^2}{2\kappa^2} - \sqrt{\lp \frac{\tilde{J}}{\kappa}(1+\sin^2\theta_0)+\frac{\gamma^2}{2\kappa^2} \rp^2 -\frac{4\tilde{J}^2}{\kappa^2}\sin^2\theta_0  }\rb^{\frac{1}{2}}
\end{align}
and is generally finite but decreases to zero as $\theta_0 \rightarrow 0$. In comparison, both s+ and a+ mode have large gaps
\begin{align}
\Omega_{s + }=& \lb \frac{2\tilde{J}}{\kappa}\cos^2\theta_0 +\frac{\gamma^2}{\kappa^2} \rb^{\frac{1}{2}} \\
\Omega_{a+}=& \lb \frac{\tilde{J}}{\kappa}(1+\sin^2\theta_0) +\frac{\gamma^2}{2\kappa^2} + \sqrt{\lp \frac{\tilde{J}}{\kappa}(1+\sin^2\theta_0)+\frac{\gamma^2}{2\kappa^2} \rp^2 -\frac{4\tilde{J}^2}{\kappa^2}\sin^2\theta_0  }\rb^{\frac{1}{2}},
\end{align}
which are dominated by the term
$\frac{\gamma^2}{2 \kappa^2} = 2 \bar{m}^2$ and are thus comparable to the Fermi energy $\epsilon_F$. Therefore, these modes are expected to be irrelevant for pairing due to an absence of separation with the Coulomb energy scale, and a corresponding lack of retardation effects.

The $s+$ branch warrants further discussion.  As alluded to above, $\mathcal{G}_{s+}$ generically exhibits poles corresponding to gapped magnon excitations.  The situation should be contrasted to $\mathcal{G}_{a-}$, whose poles indicate that the associated magnon gap decreases continuously as $\theta_0$ tends to zero---eventually becoming gapless at the $\theta_0 = 0$ spin canting transition.   Right at $\theta_0 = \pi/2$, however, a low-frequency pole suddenly emerges in $\mathcal{G}_{s+}$. This pole corresponds to the negative-frequency branch of the $s-$ mode, i.e., $\omega_{s+} = -\omega_{s-}$, meaning that $s+$ does not actually constitute a new mode.  Nevertheless, the sudden emergence of a low-frequency pole indicates that $\mathcal{G}_{s+}$ needs to be accounted for on the same footing as $\mathcal{G}_{s-}$ when calculating magnon-mediated interactions in the $\theta_0=\pi/2$ limit.  

\subsection{Magnon spectrum: order of magnitude estimates and numerics}
\label{app:order_of_magnitude_magnon}

In this Section we estimate the various scales involved in the magnon spectrum obtained from the poles of the magnon propagators $\mathcal{G}_{b\eta}(q)$ in Eq.~\eqref{Gdef} of the main text. For simplicity we first consider the critical point $\theta_0 = 0$, where the four magnon modes coalesce into only two solutions due to the restored U(1) spin rotation symmetry: a gapped ($+$) and a gapless, linearly dispersing ($-$) branch:
\begin{equation}
    \omega_{s\pm}(\vec q) = \omega_{a\pm}(\vec q) =  \sqrt{2} \bar{m} \left[ \alpha + 2 D (V + J) {\vec q}^2 \pm
  \sqrt{\alpha^2 + 4 D (V + J) {\vec q}^2 }  \right]^{1/2} ,
  \label{eq:magnon_spectrum_num}
\end{equation}
where we defined the dimensionless parameter $\alpha = \frac{V+J}{V-J}$. In  opposite limit of the SU(2) ferromagnet, with $\theta_0 = \pi/2$, there are four non-degenerate magnon modes with dispersion
\begin{align}
    \omega_{s\pm}(\vec q) &=  \sqrt{2} \bar{m}  \left[ 1 + 2 D (V + J) {\vec q}^2 \pm \sqrt{1 + 4 D (V + J) {\vec q}^2} \right]^{1/2}, \\
    \omega_{a\pm}(\vec q) &= \sqrt{2} \bar{m} \left[ \frac{V + 3 J}{V-J} +  2 D (V + J) {\vec q}^2 \pm
    \sqrt{ \frac{V + 7J}{V-J} + 4 D (V + J) {\vec q}^2 } \right]^{1/2} .
  \label{eq:magnon_spectrum_num_SU2}
\end{align}
Here the Goldstone ($s-$) mode is gapless (and quadratically dispersing), while the other three modes are gapped.

For rhombohedral multilayers (in particular following spin-orbit proximitized trilayer experiments~\cite{Patterson2025}), we can estimate $J \sim 200$ meV nm$^2$ and polarization densities (per valley) $n_0 \sim 2 \times 10^{11}$ cm$^{-2}$ relevant for the region where superconductivity is observed. The strength of intra-valley interactions $V$ (which are momentum-independent within our microscopic treatment, Eq.~\eqref{S}) can be roughly estimated from the value of the gate-screened Coulomb potential
\begin{equation}
V(|\vec q|) = \frac{q_{\mathrm{e}}^2 }{2 \epsilon_r \epsilon_0 |\vec q|} \tanh{(|\vec q| d)}
\end{equation}
at the Fermi momentum in the relevant low-density regime, $|\vec q| \sim k_F \sim \left( 0.05 - 0.1 \right) a_0^{-1}$. Here $q_e$ is the electron charge, $\epsilon_0$ the vaccum permittivity and $d$ denotes the gate distance, which is of order of a few tens of nanometers in typical experiments, such that $k_F d \gg 1$. The estimate for $V$ also depends on the amount of electronic screening in the system, which is difficult to evaluate accurately. In self-consistent Hartree-Fock treatments (e.g., \cite{Huang2023, Arp2024, Koh2024, Koh2024a}) where screening is accounted for phenomenologically by treating the relative permittivity $\epsilon_r$ as a free parameter, a best fit to the experimentally observed phase boundaries~\cite{zhou2021half, zhou2022isospin, Zhang2023, Patterson2025} is obtained in the range $\epsilon_r \sim 20-30$. With these various uncertainties in mind, we estimate $V(|\vec q| \sim k_F) \sim 1 - 3$ eV nm$^2$. The corresponding ratio of intra-valley to inter-valley interactions in our theory should then be $V/J \sim 5 - 15$.  

Using these values, the induced mass gap $\bar{m} = n_0 (J+V) \sim 2-6$ meV compares reasonably to Hartree-Fock simulations and is of the same order (although a few times larger) than the experimentally extracted peak splittings using STM in single-gated rhombohedral trilayer devices~\cite{liu2024visualizingincommensurateintervalleycoherent}. As a further consistency check, such values of $V$ (which describe a local interaction term in our theory), are of the correct order for the onset of Stoner transitions, $V D_F \approx 1$, considering typical density of states $D_F \sim 1$ eV$^{-1}$ nm$^{-2}$ in rhombohedral graphene systems near charge neutrality under a large applied displacement field.

The parameter $D$ depends sensitively on the band dispersion, including trigonal warping and corrections due to interactions, as well as the electronic density of interest. While we do not attempt a careful, quantitative calculation here (which should be carried out self-consistently), we can estimate the orders of magnitude involved. The susceptibility $\chi_1(q)$ has units of density of states, and we therefore take $\chi \sim 1$ eV$^{-1}$ nm$^{-2}$ at $|\vec q| \sim k_F \sim 0.1 a_0^{-1} \sim 0.4$ nm$^{-1}$ such that $D \sim 6$ eV$^{-1}$. 

\addELH{To reflect large uncertainties in these estimates}, we plot the magnon spectrum in Fig.~\ref{fig:app} for $D= 1$ and $10$ eV$^{-1}$, and two values of $V/J = 5, 10$, both in the ranges justified above. The separation of energy scales between the high-energy and the soft magnon modes is quite large for momentum transfers of the order $k_0 a_0 \sim 0.01 - 0.05$, which are expected to be most important for interband pairing that scatter electrons between majority and minority Fermi surfaces. (Note that $k_0$ denotes the momentum transfer between the majority and minority Fermi pockets, which can be much smaller than $k_F$ especially in the presence of trigonal warping.) 

\addELH{Using these numbers, we now estimate the scales $\Omega_0$ and $g_{\rm eff}$ appearing in our rough sketch of the critical temperature $T_c$ in the main text, Eq.~\eqref{eq:Tc}. For that we focus on the optimal canting angle $\theta_0 = 0$. From Fig.~\ref{fig:app}(a) and (b), we identify the scale $\Omega_0$ with the maximal energy of magnons within the typical momentum transfer scale shown by the shaded grey area. Taking into account the uncertainty in various parameters we take $\Omega_0 \sim 0.5 - 5$ meV. The dimensionless coupling constant
\begin{equation}
    g_{\rm eff} \sqrt{\nu_+ \nu_-} = \frac{2 \sqrt{\nu_+ \nu_-}}{D k_0^2}
\end{equation}
is more difficult to estimate as it depends on non-universal band structure quantities. Indeed, both the joint density of states of the majority and minority bands, $\sqrt{\nu_+ \nu_-}$, and the quantity $D k_0^2$ which derives from the transverse spin susceptibility, depend crucially on band structure details including trigonal warping and renormalization by interactions. In principle however, this quantity can reach the strong-coupling limit, $g_{\rm eff} \sim O(1)$, when the finite-momentum contribution to the susceptibility $D k_0^2$ is comparable to (or smaller than) the density of states.}

\begin{figure}
    \centering
    \includegraphics[width = \columnwidth]{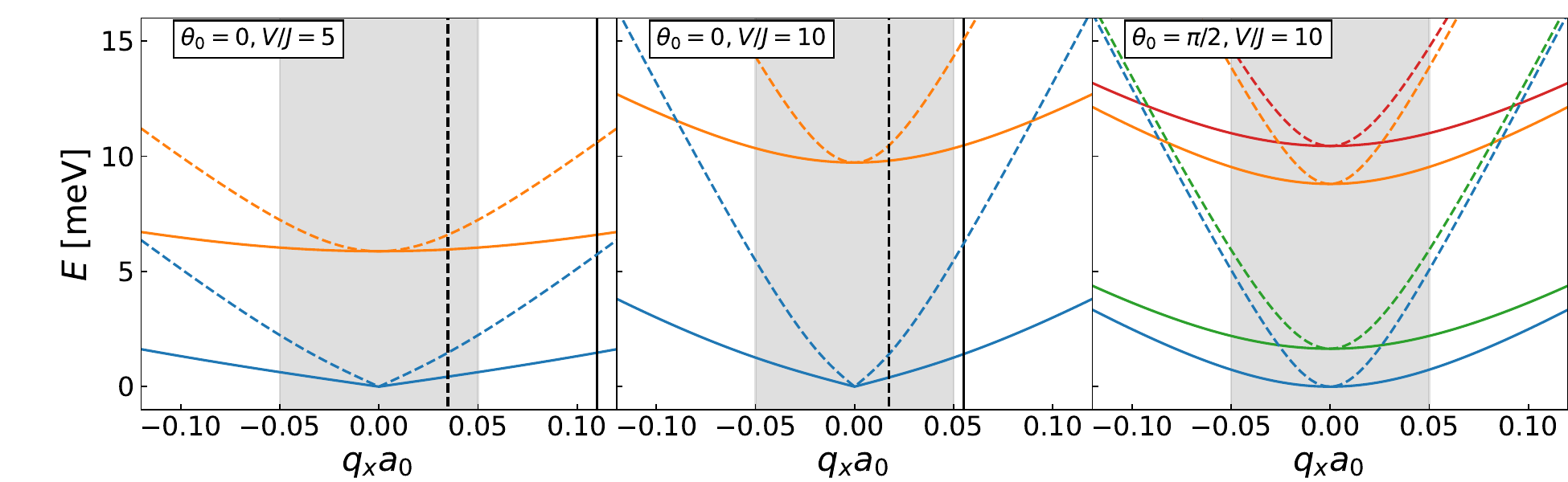}
        \caption{ \textbf{Magnon spectrum evaluated numerically from Eqs.~\eqref{eq:magnon_spectrum_num}--\eqref{eq:magnon_spectrum_num_SU2}}. We consider the critical point, $\theta_0 = 0$, for (a) $V/J = 5$ and (b) $V/J=10$, and (c) the SU(2)-symmetric ferromagnet, $\theta_0 = \pi/2$, for $V/J = 10$. The four modes are color-coded as follows: (blue = $s-$, Goldstone mode), (green = $a-$, near-critical soft mode), (orange = $s+$ and red = $a+$, high-energy modes). The characteristic scale for the minimal transferred momentum between majority and minority pockets, $k_0$, is highlighted in grey, and the value of $q_{\rm max}$ that characterizes the linear regime where the Goldstone-mode dispersion Eq.~\eqref{eq:cs} is  valid (at the critical point) is denoted by black vertical lines. Inspired by rhombohedral graphene experiments, in particular spin-orbit proximitized trilayers~\cite{Patterson2025}, we use the Hund's coupling scale $J = 200$ meV nm$^2$ and a polarization density (per valley) $n_0 = 2 \times 10^{11}$ cm$^{-2}$ throughout, for two values for $D = 1$ eV$^{-1}$ (solid lines) and $10$ eV$^{-1}$ (dashed lines) . The gaps to the two high-energy ($+$) modes are set primarily by the saddle-point value of the order parameter, $\bar{m} = n_0 (V+ J)$.
        \label{fig:app}}
\end{figure}

\section{Derivation of magnon-mediated interaction}\label{app:derive_magnon_interaction}

The spin fluctuation we considered in the last section mediates an electron-electron interaction of the following form:
\begin{align}\label{eq:magnon-mediated-interaction1}
\mathcal{S}_{I} &= \sum_{kp q} \sum_{\xi_1,\xi_2 \in\lbrace K,K'\rbrace }\sum_{\alpha\alpha'\beta\beta' =\pm } g_{\xi_1;\xi_2}^{\alpha'\alpha;\beta'\beta}(q) \overline{\Psi}_{\xi_1, k+q, \alpha'}   \overline{\Psi}_{\xi_2, p-q ,\beta'}  \Psi_{\xi_2, p,\beta}  \Psi_{\xi_1, k ,\alpha},
\end{align}
where the scattering amplitude matrix $g$ is given by
\begin{align}
g_{\xi_1,\xi_2}^{\alpha'\alpha,\beta'\beta}(q) &= -\sum_{\mu\lambda} \mathcal{M}_{\xi_1 \xi_2} ^{\mu\lambda}(q)  \sigma^{\mu}_{\alpha'
\alpha} \sigma^{\lambda}_{\beta'
\beta}  ,
\label{eq:pairing vertex 1_sup}
\end{align}
where $\mu,\lambda\in\lbrace\theta,\phi\rbrace$, $\sigma^{\theta} = \lp
\begin{matrix}
    0 & 1\\
    1 & 0
\end{matrix}\rp $, $\sigma^{\phi} = \lp
\begin{matrix}
    0 & -i\\
    i & 0
\end{matrix}\rp $, whereas $\mathcal{M}_{\xi_1 \xi_2}$ is defined as
\begin{align}
\mathcal{M}_{\xi_1 \xi_2} 
&= \left\langle  
\lp
\begin{matrix} 
\delta m^\theta_{\xi_1} \delta m^\theta_{\xi_2} & \delta m^\theta_{\xi_1} \delta m^\phi_{\xi_2} \\
\delta m^\phi_{\xi_1} \delta m^\theta_{\xi_2} & \delta m^\phi_{\xi_1} \delta m^\phi_{\xi_2} 
\end{matrix}
\rp
\right\rangle 
\end{align}
Using definition of symmetric and antisymmetric modes in Eq.~\eqref{eq:def_ms_ma}, we express $\mathcal{M}$ in terms of symmetric and antisymmetric modes' contributions: 
\begin{align}
\mathcal{M}_{\xi_1 \xi_2} &= \frac{1}{2}\langle \delta M_s \delta M_s\rangle + \frac{1}{2}\tau_{\xi_1}\tau_{\xi_2}\langle \delta M_a \delta M_a\rangle  
\end{align}
Both terms can be read off from Eq.~\eqref{SmagFinal}  by taking the inverse of the 2-by-2 matrix in Eq.~\eqref{SmagFinal}:
\begin{align}
\mathcal{M}_{\xi_1 \xi_2}= \frac{1}{2} \frac{1}{Q^\theta_s Q^\phi_s+\gamma^2\nu^2} 
\lp \begin{matrix}
Q^\phi_s & -\gamma\nu \\
+\gamma\nu & Q^\theta_s 
\end{matrix}
\rp 
+ \frac{1}{2}\tau_{\xi_1}\tau_{\xi_2} \frac{1}{Q^\theta_a Q^\phi_a+\gamma^2\nu^2}
\lp \begin{matrix}
Q^\phi_a & -\gamma\nu \\
+\gamma\nu & Q^\theta_a 
\end{matrix}
\rp
\end{align}
where $\tau_{\xi}= \pm1$ for $\xi=K,K'$. So $\tau_{\xi_1}\tau_{\xi_2}=\pm1$ represents whether $\xi_1$ and $\xi_2$ are the same valley indices. This factor enters the second term because the second term represents the contribution of the antisymmetric magnon modes which feature opposite amplitudes in valleys $K$ and $K'$.
Next, we explicitly compute all the magnon-mediated scattering amplitudes
\begin{align}
g^{+-,+-}_{\xi_1, \xi_2}(q)  &= -\frac{1}{2}\frac{Q^\phi_s+(-i)(-\gamma\nu)+(-i)(\gamma\nu) + (-i)^2 Q^\theta_s}{Q^\theta_s Q^\phi_s+\gamma^2\nu^2}
-\tau_{\xi_1} \tau_{\xi_2} \frac{1}{2}\frac{Q^\phi_a+(-i)(-\gamma\nu)+(-i)(\gamma\nu) + (-i)^2 Q^\theta_a}{Q^\theta_a Q^\phi_a+\gamma^2\nu^2},\nonumber\\
&=-\frac{1}{2}\frac{Q^\phi_s- Q^\theta_s}{Q^\theta_s Q^\phi_s+\gamma^2\nu^2}
-\tau_{\xi_1} \tau_{\xi_2} \frac{1}{2}\frac{Q^\phi_a- Q^\theta_a}{Q^\theta_a Q^\phi_a+\gamma^2\nu^2},
\label{eq:g^+-,+-_sup}
\\
g^{-+,+-}_{\xi_1, \xi_2}(q)  &= - \frac{1}{2}\frac{Q^\phi_s+(-i)(-\gamma\nu)+i(\gamma\nu) + i(-i) Q^\theta_s}{Q^\theta_s Q^\phi_s+\gamma^2\nu^2}
-\tau_{\xi_1} \tau_{\xi_2} \frac{1}{2}\frac{Q^\phi_a+(-i)(-\gamma\nu)+i(\gamma\nu) + i(-i) Q^\theta_a}{Q^\theta_a Q^\phi_a+\gamma^2\nu^2},\nonumber\\
&= -\frac{1}{2}\frac{Q^\phi_s + Q^\theta_s + 2i\gamma\nu }{Q^\theta_s Q^\phi_s+\gamma^2\nu^2}
-\tau_{\xi_1} \tau_{\xi_2} \frac{1}{2}\frac{Q^\phi_a + Q^\theta_a + 2i\gamma\nu}{Q^\theta_a Q^\phi_a+\gamma^2\nu^2}
\label{eq:g^-+,+-_sup}
\end{align}
where
\begin{align}
&Q^{\theta}_s = 
2\tilde{J}\cos^2\theta_0+\kappa \nu^2 +D {\vec q}^2, \quad Q^{\phi}_s = 
\kappa \nu^2 +D {\vec q}^2, \label{eq:Qs}\\
&Q^{\theta}_a = 
2\tilde{J}\sin^2\theta_0+\kappa \nu^2 +D {\vec q}^2, \quad Q^{\phi}_a = 
2\tilde{J} + \kappa \nu^2 +D {\vec q}^2 .\label{eq:Qa}
\end{align}

As a consistency check, we look at the case of canting critical point where $\theta_0=0$.  Eqs.~\eqref{eq:g^+-,+-_sup} and \eqref{eq:g^-+,+-_sup} show the intravalley $g^{-+,-+}_{\xi,\xi}$ and intervalley $g^{-+,+-}_{K,K'}$ amplitudes vanish, in line with expectations from symmetry: At $\theta_0 = 0$, the system preserves global U(1) spin rotation symmetry around the $z$ axis, which in our rotated basis enforces conservation of $S^z_{\rm tot} = \sum_\xi\int_k \overline{\Psi}^\dagger_{\xi,k}\tau_\xi\sigma^z \Psi_{\xi,k}$. Intravalley $g^{-+,-+}_{\xi,\xi}$ amplitudes encode processes that change $S^z_{\rm tot}$ by flipping two spins in a given valley; intervalley $g^{-+,+-}_{K,K'}$ amplitudes similarly violate this conservation law.

Next, plugging Eq.~\eqref{eq:Qs} and Eq.~\eqref{eq:Qa} into Eq.~\eqref{eq:g^+-,+-_sup} and  Eq.~\eqref{eq:g^-+,+-_sup}, and dropping higher-order terms in $q$ and $\nu$ yields
\begin{align}
g^{+-,+-}_{\xi_1,\xi_2}(q)  \sim& \frac{\tilde{J} \cos^2 \theta_0}{\lb 2\tilde{J}\kappa \cos^2\theta_0 +\gamma^2 \rb \nu^2 + 2\tilde{J}\kappa \cos^2\theta_0 D {\vec q}^2} \nonumber \\
&-\tau_{\xi_1} \tau_{\xi_2} \frac{\tilde{J} \cos^2 \theta_0}{4\tilde{J}^2\sin^2\theta_0 + \lb 2\tilde{J}\kappa \lp 1+\sin^2\theta_0\rp +\gamma^2 \rb \nu^2 + 2\tilde{J}\kappa (1+\sin^2\theta_0) D {\vec q}^2},\label{eq:g+-+-}\\
g^{-+,+-}_{\xi_1,\xi_2}(q)  \sim&  -\frac{\tilde{J}\cos^2\theta_0 + \kappa \nu^2 + D {\vec q}^2 + i\gamma \nu}{\lb 2\tilde{J}\kappa \cos^2\theta_0 +\gamma^2 \rb \nu^2 + 2\tilde{J}\kappa \cos^2\theta_0 D {\vec q}^2} \nonumber
\\
&-\tau_{\xi_1} \tau_{\xi_2} \frac{\tilde{J}(1+\sin^2\theta_0) + \kappa \nu^2 + D {\vec q}^2 + i\gamma \nu}{4\tilde{J}^2\sin^2\theta_0 + \lb 2\tilde{J}\kappa \lp 1+\sin^2\theta_0\rp +\gamma^2 \rb \nu^2 + 2\tilde{J}\kappa (1+\sin^2\theta_0) D {\vec q}^2},\label{eq:g+--+}
\end{align}

For generic canting $0<\theta_0<\pi/2$ we have
\begin{align}
g^{+-,+-}_{\xi_1,\xi_2}(q) \sim & \frac{ \cos^2\theta_0 }{2z_s \kappa } \frac1{ \nu^2 +c_s^2  {\vec q}^2} -\tau_{\xi_1} \tau_{\xi_2} \frac{ \cos^2\theta_0 }{2z_a \kappa } \frac1{ \Omega_{a-}^2 + \nu^2 +c_a^2  {\vec q}^2},\\
g^{-+,+-}_{\xi_1,\xi_2}(q) \sim & \frac{ \cos^2\theta_0 }{2z_s \kappa } \frac1{ \nu^2 +c_s^2  {\vec q}^2} -\tau_{\xi_1} \tau_{\xi_2} \frac{ 1+\sin^2\theta_0 }{2z_a \kappa } \frac1{ \Omega_{a-}^2 + \nu^2 +c_a^2  {\vec q}^2},
\end{align}
Here, we have defined
\begin{align}
c_s &= \lp \frac{D}{\kappa z_s} \rp^{1/2} \cos\theta_0, \quad z_s = \cos^2\theta_0 +\frac{V-J}{2 J }, \\
c_a &= \lp \frac{D}{\kappa z_a} \rp^{1/2} (1+\sin^2\theta_0)^{1/2}, \quad z_a =(1+\sin^2\theta_0) +\frac{V-J}{2 J } \\
\Omega_{a-} &=  \sqrt{\frac{2\tilde{J}}{ z_a \kappa}}\sin\theta_0 \sim \lp \frac{J\sin\theta_0}{V}\rp \sqrt{\frac{1}{V\kappa }}\sim \lp \frac{J\sin\theta_0}{V}\rp \epsilon_F,\quad \text{when $V\gg J$} .
\end{align}
Here we have used the identity $\gamma^2/\tilde{J} \kappa = (V-J)/J$, which follows from Eq.~\eqref{eq:parameters}, and assumed that $\bar{m} \sim \epsilon_F$ as in the main text.

For the case of $\theta_0=\pi/2$ which is achieved in the absence of Ising SOC, we have
\begin{align}
g^{+-,+-}_{\xi_1,\xi_2}(q) =& 0\label{eq:g+-+-,theta=pi/2}\\
g^{-+,+-}_{\xi_1,\xi_2}(q) \sim & -\frac{1}{\kappa \nu^2 + D {\vec q}^2 - i\gamma \nu}  -\tau_{\xi_1} \tau_{\xi_2} \frac{ 1}{2z_a \kappa } \frac1{ \Omega_{a-}^2 + \nu^2 +c_a^2  {\vec q}^2} . \label{eq:g-++-,theta=pi/2}
\end{align}
Here, in $g^{-+,+-}$, the first term describes the contribution of symmetric magnon modes, which merge into one branch of quadratic magnon at $\theta_0 = 0$. The second term describes the contribution of the gapped antisymmetric modes. At small $q$ and $\nu$, Eq.~\eqref{eq:g-++-,theta=pi/2} reduces to $g^{-+,+-}_{\xi_1,\xi_2}(q) \sim -\frac{1}{-i \gamma \nu + D \vec{q}^2}$.

\addELH{\section{Magnon spectrum in the presence of an in-plane magnetic field $B_\parallel$}} \label{app:B-dependent_magnon_gap}

\addELH{Here we study the magnon spectrum in the presence of an in-plane magnetic field $B_\parallel$. Because $B_\parallel$ explicitly breaks the $U(1)$ spin rotation symmetry around the Ising axis, we expect the Goldstone and critical modes to be gapped out. As we shall see, the induced gap for the Goldstone mode scales as $\Delta_{\rm mag} \sim |B_\parallel| $ in the absence of spin-orbit coupling ($\theta_0 \rightarrow \pi/2$), but as $\Delta_{\rm mag} \sim \sqrt{|B_\parallel|}$ within the canted phase that arises when $\lambda_I$ is introduced. The magnon gap is parametrically enhanced at the critical point ($\theta_0 \rightarrow 0$), where $\Delta_{\rm mag} \sim |B_\parallel|^{1/3}$.} 

We return to the action in Eq.~\eqref{S3} and include the Zeeman coupling of electrons to the applied magnetic field, which we take along the $x$ axis for concreteness, ${\vec B_\parallel} = - B_\parallel \hat{\vec{x}}$:
\begin{align}
\mathcal{S} &= \sum_{\xi}\int_k\overline{\psi}_{\xi, k} \lp -i\omega -\mu + \epsilon_{\xi,\vec{k}} - \frac{\lambda_I}{2}\tau_{\xi}\sigma^z  - B_\parallel \sigma^x \rp \psi_{\xi, k} +\int_q\bigg{[}\frac{1}{2}\sum_{\xi,\xi'}g^{-1}_{\xi\xi'}\vec m_\xi (q) \cdot \vec m_{\xi'}(-q) -\sum_{\xi} \vec m_\xi (q) \cdot \vec s_{\xi}(-q)\bigg{]},
\label{eq:action_appendix_Bfield}
\end{align}
where we set the Bohr magneton $\mu_B = 1$ for notational simplicity. We now define 
\begin{equation}
    \vec{m}_\xi(q) \rightarrow \delta(q) \left( -\frac{\lambda_I}{2} \tau_{\xi}{\bf \hat{z}} - B_\parallel {\bf \hat{x}} + \bar{\vec{m}}_\xi \right),
    \label{eq:definition_m_Bfield}
\end{equation}
such that $\bar{\vec{m}}_\xi$ is the net effective field experienced by electrons in valley $\xi$. 

An in-plane field $B_\parallel$ remains compatible with the canted, valley-balanced solution introduced in the main text (as opposed to, e.g., an out-of-plane magnetic field, which would induce valley imbalance). We are thus looking for a spin-canted saddle-point solution of Eq.~\eqref{eq:action_appendix_Bfield} with the same form as Eq.~\eqref{eq:saddle_point_solution_ansatz},
\begin{equation}
    \bar{\vec{m}}_\xi = \bar{m}(\sin\theta {\bf \hat{x}} + \tau_{\xi} \cos\theta{\bf \hat{z}}).
    \label{eq:definition_m_appendix}
\end{equation}
Integrating out the fermions and dropping constants independent of $\bar{m}$ and $\theta$, we obtain an analog to Eq.~\eqref{eq:free_energy_density_angle} that now also incorporates $B_\parallel$,
\be
F = \mathcal{A}(\bar{m}) +  \frac{\bar{m}^2}{V+J} + \frac{2 J \bar{m}^2 \cos^2\theta}{ V^2-J^2}   - \frac{\lambda_I \bar{m}  \cos\theta}{V-J } -\frac{ 2 B_\parallel \bar{m} \sin\theta}{V+J} .
\label{eq:free_energy_density_angle_appendix}
\ee
The first term on the right-hand side represents the contribution from fermions in two bands split in energy by $\pm \bar{m}$ \addELH{(where $\bar{m}$ includes magnetic field, spin-orbit, and interaction effects as defined in Eq.~\eqref{eq:definition_m_Bfield})}. Minimizing with respect to $\theta$, the saddle-point condition reads
\begin{equation}\label{eq:canting angle condition}
    \frac{-4 J \bar{m} \cos\theta \sin \theta}{ V^2-J^2} + \frac{\lambda_I  \sin \theta}{V-J } - \frac{ 2 B_\parallel \cos\theta}{V+J}  = 0.
\end{equation}

\subsection{Generic canting angle regime}

Unfortunately, this expression does not admit a closed-form solution for $\theta$ in the presence of $B_\parallel$. To make analytical progress, we rewrite $\theta$ as a $B_\parallel$-induced deviation on top of the solution at $B_\parallel=0$, i.e. $\theta = \theta_0+ \delta \theta$ with $\theta_0 = \arccos[\lambda_I (V+J)/4 J \bar{m}_0]$, and expand to leading order in the deviation $\delta \theta$. \addELH{We also expand $\bar{m} = \bar{m}_0 + \chi_B B_\parallel$ to capture the leading-order modification to the effective field. These expansions are well-defined if the parameters $\theta$ and $\bar{m}$ are smoothly varying with respect to $B_\parallel$; as we shall see below, this assumption breaks down at the critical point $\theta_0 = 0$, which requires a separate treatment.}
The canting angle condition Eq.~\eqref{eq:canting angle condition} then yields an expression for $\delta \theta$ in terms of the applied magnetic field, which to leading order in $B_\parallel$ reads
\begin{equation}
    \delta \theta =  \frac{2}{V+J} \frac{B_\parallel \lambda_I}{(\lambda_I^c)^2 - \lambda_I^2} \left( V - J + \addELH{2 J \chi_B \sqrt{1 - \frac{\lambda_I^2}{(\lambda_I^c)^2}} } \right),
   \label{eq:canting_B_condition}
\end{equation}
where 
\begin{equation}
  \lambda_I^c = \frac{4 J\bar{m}_0}{V+J}    
\end{equation}
is the critical Ising value where the canted state transitions to the spin-valley-locked state with $\theta_0 = 0$. The divergence of $\delta \theta$ when $\lambda_I \rightarrow \lambda_I^c$
reflects the breakdown of our expansion above and the diverging susceptibility to a gap opening at the phase transition.

As in the main text, we use the saddle-point condition $|{\vec s}_\xi(q)| = \delta(q) n_0$---obtained by minimizing the action Eq.~\eqref{eq:action_appendix_Bfield} with respect to $\vec m_\xi (q)$---to derive a condition between the spin polarization density $n_0$ and the magnitude $\bar{m}$ of the effective field that does not involve the $\mathcal{A}(\bar{m})$ contribution:
\begin{align} \label{eq:n_0 with B generic}
    n_0 = \left|\frac{\bar{m} \sin\theta - B_\parallel}{V+J}{\bf\hat{x}} + \frac{\bar{m}\cos\theta-\lambda_I/2}{V-J}{\bf \hat{z}}\right|.
\end{align}
We expand the right-hand side to linear order in small angle and magnitude deviations from the zero-field solution, $\theta = \theta_0 + \delta \theta$ and $\bar{m} = \bar{m}_0 + \chi_B B_\parallel$, and use Eq.~\eqref{eq:canting_B_condition} for $\delta \theta$.  To linear order in $B_\parallel$, we obtain
\begin{equation} \label{eq:n_0 with B}
n_0 = \frac{\bar{m}}{V+J} - \frac{B_\parallel \lambda_I^c}{(V+J) \sqrt{(\lambda_I^c)^2 - \lambda_I^2}}.
\end{equation}

We now examine the impact of this result on the magnon dispersion. The coefficients appearing in the spin susceptibilities---see Eq.~\eqref{eq:parameters}---are altered: for example, $C=\frac{n_0}{\bar{m}}$ is no longer equal to $\frac{1}{V+J}$, but instead 
\be
C = \frac{1}{V+J} - C_B, \quad C_B = \frac{B_\parallel}{\bar{m}_0 (V+J)} \frac{\lambda_I^c}{\sqrt{(\lambda_I^c)^2 - \lambda_I^2}}.
\label{Ceq}
\ee
This result qualitatively impacts the Goldstone mode dispersion because $C_B$---which we observe does not depend on the undetermined parameter $\chi_B$---enters the diagonal elements of the magnetic correlator $Q_s^\phi$; see Eqs.~\eqref{SmagFinal} and Eq.~\eqref{eq:Q main text}. As a reminder, in the absence of $B_\parallel$, we found $Q^\phi_s = \frac{1}{V+J} - \frac{n_0}{\bar m} + \kappa \nu^2 + D {\vec q}^2 = \kappa \nu^2 + D {\vec q}^2 $. In the presence of $B_\parallel$, as shown in Eq.~\eqref{eq:n_0 with B} the constant term  $\frac{1}{V+J} - \frac{n_0}{\bar m} = C_B$ no longer vanishes. Accounting for this correction to $Q^\phi_s$ and solving the pole of Eq.~\eqref{eq:g_eigenvalue_appendix},
we find that the Goldstone magnon mode obtains a gap
\be
\Delta_{\rm{mag}} = \sqrt{ \left(\frac{C_B}{\zeta_0} \right) + O(B_\parallel^2)} \approx \sqrt{ \frac{B_\parallel}{\zeta \bar{m}_0 (V+J)}} \sqrt{\frac{\lambda_I^c}{\sqrt{(\lambda_I^c)^2 - \lambda_I^2}}}, \quad \zeta = \kappa + \frac{\gamma^2}{2\tilde{J} \cos^2 \theta_0}.
\label{eq:magnon_gap_result}
 \ee
Here we ignored $B_\parallel$-induced corrections to the other diagonal correlators $Q^\theta_{s/a}$ and $Q^\phi_a$, as well as parameters $\kappa$, $D$, $\gamma$. These corrections are all subleading because these quantities are originally non-vanishing at $q=0$---an $O(B_\parallel)$ correction to them would give subleading corrections that can be absorbed into the $O(B_\parallel^2)$ term in Eq.~\eqref{eq:magnon_gap_result}.

For an order of magnitude estimate, we use the parameter values in Appendix \ref{app:order_of_magnitude_magnon}: $V \sim 2$ eV nm$^2$, $J \sim 200$ meV nm$^2$, $\bar{m} \sim 4$ meV, yielding $\lambda_I^c = 1.5$ meV. Re-introducing the Bohr magneton and taking a typical value $\lambda_I \sim 1$ meV, we obtain
\be
\Delta_{\rm{mag}}  \sim \sqrt{ \frac{\mu _B B_\parallel}{ \left( \frac{1}{16 \text{meV}} + \frac{1}{8 \text{meV}} \right)}} \sim \sqrt{ 5 \text{meV} \times \mu _B B_\parallel}.
 \ee
For an in-plane field in the $10 - 100 \text{mT}$ range, we expect $\Delta_{\rm mag} \sim \left( 0.05 - 0.2 \right)~\text{meV}$. This gap is likely too small to meaningfully affect pairing \addELH{mediated by finite-momentum magnons} 
which, as discussed in Appendix \ref{app:order_of_magnitude_magnon}, is controlled by an energy scale $\Omega_0 \sim 1$ meV, except close to the canting transition where the magnon gap is enhanced.  \addELH{These energy gaps are however well-suited for spectroscopic study using microwave photons, for example using resonator architectures inspired by circuit QED~\cite{Zhang2014magnons, Tabuchi2014magnons, Boettcher2024}.}

\subsection{Magnon gap at the critical point}
 
The magnon gap in Eq.~\eqref{eq:magnon_gap_result} is obtained in the generic canting regime where $\theta_0$ approaches neither $0$ nor $\pi/2$. As noted above, when $\theta_0 \rightarrow 0$, our expansion fails due to diverging susceptibility to a gap opening at the canting phase transition.
Right at the spin-canting transition, a similar analysis yields a parametrically larger magnon gap $\Delta_{\rm mag} \sim B_{\parallel}^{1/3}$ for both the $s-$ and $a-$ modes.

To derive this result, we now specialize to $\lambda_I = \lambda_I^c$ and write the free energy in Eq.~\eqref{eq:free_energy_density_angle_appendix} as
\be
F(\bar m ,\theta) = F_0(\bar m) - \frac{2 J \bar{m}^2 \sin^2\theta}{ V^2-J^2}   - \frac{\lambda_I^c \bar{m}  (\cos\theta-1)}{V-J } -\frac{ 2 B_\parallel \bar{m} \sin\theta}{V+J} .
\label{F_rewriting}
\ee
At $B_\parallel = 0$, the optimal angle is $\theta_0 = 0$, and hence in this rewriting all the free energy is incorporated through $F_0$.  
Moreover, in this limit $\bar m_0$ minimizes the free energy, implying that $F_0$ should take the form (up to a constant) 
\begin{equation}
  F_0 (\bar m) = \frac{\alpha}{2} (\bar m-\bar m_0)^2  
  \label{F0}
\end{equation}
for some phenomenological parameter $\alpha > 0$.  A non-zero $B_\parallel$ \emph{explicitly} enters the free energy function only through the last term in Eq.~\eqref{F_rewriting} (though when minimized, all terms will be impacted through implicit $B_\parallel$ dependence in the optimal $\bar m$ and $\theta$ values); Eq.~\eqref{F0} thus continues to hold even with $B_\parallel \neq 0$.  

The saddle-point conditions $\partial F/\partial\theta = 0$ and $\partial F/\partial \bar m = 0$ yield the following constraints:
\begin{align}
    0 &=\frac{\lambda_I^c  \sin\theta}{V-J}\left(1-\frac{\bar m}{\bar m_0}\cos\theta\right)- \frac{2 B_\parallel \cos\theta}{V+J}
    \\
    0 &= \alpha (\bar m - \bar m_0)  + \frac{\lambda_I^{c}(1-\cos\theta)}{V-J}\left[1- \frac{\bar m}{\bar m_0}(1+\cos\theta)\right]  - \frac{2B_\parallel\sin\theta }{V+J}.
\end{align}
Writing $\bar m = \bar m_0 + \delta m$ and anticipating that $\delta m$ and $\theta$ both scale to zero as a power-law (with exponent less than 1) when $B_\parallel \rightarrow 0$, we expand the above equations for small $B_\parallel$ to arrive at
\begin{align}
    \frac{\lambda_I^c \theta}{V-J}\left(\frac{1}{2}\theta^2-\frac{\delta m}{\bar m_0}\right)&\approx \frac{2 B_\parallel}{V+J}
    \label{smallB1}
    \\
    \alpha \delta m  - \frac{\lambda_I^{c}\theta^2}{2(V-J)}  &\approx 0.
    \label{smallB2}
\end{align}
Equations~\eqref{smallB1} and \eqref{smallB2} are solved with
\begin{equation}\label{eq:scaling}
  \theta \sim \beta_\theta |B_\parallel|^{1/3} ,~~~~\delta m \sim \beta_m |B_\parallel|^{2/3}, \quad \beta_\theta = \lp \frac{4(V-J)}{\lambda_I^c(V+J)(1 - \frac{\lambda_I^c }{\alpha \bar{m}_0(V-J) })} \rp^{1/3}, \quad \beta_m = \frac{\lambda_I^c}{2(V-J)\alpha }\beta_\theta ^2  .
\end{equation}

The prefactors $\beta_\theta$ and $\beta_m$ are both positive for physically relevant parameters.
To see positivity of $\beta_\theta$, we note that $\alpha$ is the longitudinal stiffness and should be set by Fermi-surface parameters. Since $\alpha$ has dimensions of density of states $\nu_0$, we can estimate it as $\alpha\sim 1/V$ (assuming $\nu_0 V\lesssim 1$). Therefore, by counting powers in $J/V$,  we have $\frac{\lambda_I^c}{\alpha \bar{m}_0(V-J)} = \frac{4J}{(V^2-J^2)\alpha}\sim O(J/V)\ll 1$---indeed ensuring $\beta_\theta>0$.

These scaling laws allow us to determine the $B_\parallel$ dependence of the magnon gap. Plugging Eq.~\eqref{eq:scaling} into Eq.~\eqref{eq:n_0 with B generic} yields $\delta n \sim B^{2/3}$, where $\delta n$ denotes the $B_\parallel$-induced change of $n_0$. Next, we look into the quantity $C_B$, the change in $n_0/\bar{m}$ due to $B_\parallel$ [recall left side of Eq.~\eqref{Ceq}], which directly enters the magnon gap:
$
C_B = 
\frac{\delta n}{\bar m_0} - \frac{n_0 \delta m}{\bar m_0^2}
\sim B_\parallel^{2/3}$.
Finally, plugging the latter scaling into the first equality of Eq.~\eqref{eq:magnon_gap_result} reveals that the magnon gap has a singular dependence on $B_\parallel$ as claimed earlier: $\Delta_{\rm mag}\sim B_\parallel^{1/3}$.

\subsection{Ferromagnetic limit, $\theta_0 \rightarrow \pi/2$}

Unlike for a generic finite $\theta_0$ where one only needs to account for the impact of $B_\parallel$ on $Q_s^{\theta}$, in the case of $\theta_0=\pi/2$ we need to consider
$Q_s^{\theta}$ and $Q_s^{\phi}$ as both symmetric modes are gapless. Namely, we rewrite $Q_s^{\phi}=Q_s^{\theta} = C_B + \kappa\nu^2+D\vec q^2$.
This results in a magnon gap of $\Delta_{\text{mag}} = |C_B| /\gamma \sim O(B_\parallel)$.

\subsection{Effect on pairing strength and critical temperature}

\addELH{We now consider the effect of field-induced magnon gaps on the critical temperature $T_c$ for superconductivity. We first consider  the generic canting-angle regime, where $\theta_0$ does not approach $0$ or $\pi/2$, and where the low-energy physics is dominated by the Golstdone magnon.}

\addELH{Above we showed that the Goldstone magnon acquires a gap $\Delta_{\rm mag}$ in an in-plane magnetic field. The small-${\vec q}$ dispersion becomes
\begin{equation}
    \omega_{s-}(\vec q) = \sqrt{ c_s^2 {\vec q}^2 + \Delta^2_{\rm mag} },
\end{equation}
where for simplicity we assumed that the magnon velocity $c_s$ is not renormalized by $B_\parallel$: the only modification to the Goldstone-mode spectrum comes through the gap.} The Goldstone contribution to the pairing strength is then modified to
\begin{equation}\label{eq:geff_appendix}
    g_{\rm eff} = \frac{2 J \cos^2 \theta_0}{V \kappa} \frac{1}{c_s^2 k_0^2 + \Delta^2_{\rm mag}}
\end{equation}
Replacing $\Omega_0 \rightarrow \sqrt{c_s^2 k_0^2 +  \Delta_{\rm mag}^2}$ in Eq.~\eqref{eq:Tc} (and focusing only on the Goldstone contribution), our estimate for the critical temperature becomes
\begin{equation}
    T_c \approx \sqrt{c_s^2 k_0^2 + \Delta_{\rm mag}^2} \exp \left(- \frac{V \kappa (c_s^2 k_0^2 + \Delta^2_{\rm mag})}{2 J \cos^2 \theta_0 \nu_{\rm eff}} \right).
\end{equation}
Expanding this exponential for small $\Delta_{\rm mag}^2 \sim |B_\parallel|$, and using the 
result Eq.~\eqref{eq:magnon_gap_result} for the magnon gap, we find
\begin{equation}
    T_c - T_c^{(0)} \approx  \frac{|B_\parallel|}{\zeta \bar{m} (V+J)} \frac{\lambda_I^c}{\sqrt{(\lambda_I^c)^2 - \lambda_I^2}} \frac{1}{2 c_s k_0}  \left( 1 - \frac{2}{g_{\rm eff}^{(0)} \nu_{\rm eff}} \right) \exp \left( - \frac{1}{g^{(0)}_{\rm eff} \nu_{\rm eff}} \right)
\end{equation}
where $T_c^{(0)}$ and $g_{\rm eff}^{(0)} = \cos^2 \theta_0/(Dk_0^2)$ are respectively the critical temperature and the pairing strength in the absence of magnetic field.

This analysis shows that the critical temperature is modified linearly with the in-plane field---in contrast, if the normal state preserved time-reversal symmetry, then  $T_c$ would acquire a $B_\parallel^2$ shift through the usual Ginzburg-Landau arguments. The $\sim |B_\parallel|$ change of  $T_c$ at low fields thus constitutes a further experimental prediction of our scenario. Within our simplified treatment above, we find that $T_c$ is reduced by applying an in-plane field, provided that the superconductor is not too strongly coupled, $g_{\rm eff}^{(0)} \nu_{\rm eff} < 2$. In the strongly coupled  limit, interestingly we find that $T_c$
initially \emph{increases} with $B_\parallel$, until the higher-order contributions from the exponential suppress pairing. It would be interesting to verify whether this phenomenology holds in more realistic modeling of magnon-induced depairing in rhombohedral graphene multilayers.

At the canting transition, the critical $s-$ and $a-$ modes are degenerate and acquire a gap $\Delta_{\rm mag} \sim |B_\parallel|^{1/3}$ in an in-plane magnetic field. The corresponding pairing interaction, Eq.~\eqref{eq:g_eff theta=0}, becomes
\begin{equation}
    g_{\rm eff} = \frac{4J}{V \kappa} \frac{1}{c^2 k_0^2 + \Delta_{\rm mag}^2}
\end{equation}
where $c = c_s = c_a = \sqrt{\frac{2JD}{V \kappa}}$ is the (degenerate) magnon velocity at the critical point. \addELH{Following the same steps as above, 
%
and expanding to leading order in $|B_\parallel|$ yields a temperature change that scales as
\begin{equation}
    T_c - T_c^{(0)} \sim |B_\parallel|^{2/3}.
\end{equation}
}

\section{Landau damping of magnons}
\label{sec:landau_damping}

\addELH{In this Appendix we derive the Landau damping of the soft magnon modes considered in this work. Focusing on a single valley and dropping the index $\xi$ for simplicity,} we consider a spin-split two-band model with dispersion
\begin{align}
    \epsilon_{{\vec k}}^\pm &= \frac{k^2}{2m_0} \mp \bar{m},
\end{align}
where $\bar{m} > 0$ is the band splitting induced by \addELH{a combination of Ising spin-orbit coupling and interaction-induced exchange splitting.}  The chemical potential $\mu$ is common to both spin species. The Fermi momenta are
\begin{align}
    k_F^{\pm} &= \sqrt{2m_0(\mu \pm \bar{m})}.
\end{align}

We study the damping of a magnon of momentum ${\vec q}$ and real frequency $\omega$, which corresponds to spin-flip particle-hole excitations between the majority and minority bands. The damping rate $\Gamma$ is given by the imaginary part of the transverse spin susceptibility $\Pi^{+-}$. \addELH{From Eqs.~\eqref{eq:chi_1_result} and \eqref{eq:chi_2_result}, we first recall the Matsubara frequency expression
\begin{align}
    \Pi^{+-}(i\nu , {\vec q}) = - T \sum_{\omega'} \int \frac{d^2{\vec k}}{(2\pi)^2} G_+(\omega' + \nu,\vec k+\vec q)G_-(\omega',\vec k) = \int \frac{d^2{\vec k}}{(2\pi)^2} \frac{f(\epsilon_{{\vec k}+{\vec q}}^+)-f(\epsilon_{{\vec k}}^-) }{i \nu + (\epsilon_{{\vec k}+{\vec q}}^+- \epsilon_{{\vec k}}^-)  }.
\end{align}
Analytically continuing to real frequency using $i \nu \rightarrow \omega + i 0^+$, we obtain the damping rate
}

\begin{align}
    \Gamma(\omega, {\vec q}) = \Im \Pi^{+-}(\omega+i0^+,{\vec q})
    =
    -\pi \int \frac{d^2{\vec k}}{(2\pi)^2}
    \, 
\lb f(\epsilon_{{\vec k}+{\vec q}}^+)-f(\epsilon_{{\vec k}}^-)\rb
    \delta \left( \omega -(\epsilon_{{\vec k}+{\vec q}}^+ -\epsilon_{{\vec k}}^-) \right).
    \label{eq:gamma}
\end{align}

We focus on the regime $|{\vec q}| > q_c \equiv k_F^+ - k_F^-$ \addELH{(i.e., where the magnon momentum is above the threshold for interband transitions)} and $\omega \ll \mu$ \addELH{(which is required  in order to benefit from the Anderson-Morel or retardation effect)}.
We align ${\vec q}$ along the $x$-axis and write ${\vec k} = k(\cos\theta, \sin\theta)$. The energy difference in the delta function becomes
\begin{align}
    \omega - (\epsilon_{{\vec k}+{\vec q}}^+ - \epsilon_{{\vec k}}^-)
    &= \omega + \frac{{\vec k}^2}{2m_0} - \frac{({\vec k}+{\vec q})^2}{2m_0} + 2\bar{m} \\
    &= \omega - \frac{2kq\cos\theta + q^2}{2m_0} + 2\bar{m} .
\end{align}
The delta function then reads
\begin{align}
    \delta\left( \omega - \frac{2kq\cos\theta + q^2}{2m_0} + 2\bar{m} \right)
    = \frac{m_0}{q} \delta\left(k_x - k_0\right),\label{eq:damping delta function}
\end{align}
\addELH{where we defined the momentum scale
\be 
k_0 = \frac{m_0}{q} \left( \omega -\frac{q^2}{2m_0} + 2\bar{m} \right),
\label{eq:damping k0}
\ee
which depends on the external frequency $\omega$ and momentum magnitude $q$, but importantly not on its direction. This momentum scale defines a particular value of $k_x$ which contributes to the damping---see the red line in Fig.~\ref{fig:damping integral}.}
Inserting into Eq.~\eqref{eq:gamma}, one finds
\begin{figure}
    \centering
    \includegraphics[width=0.5\linewidth]{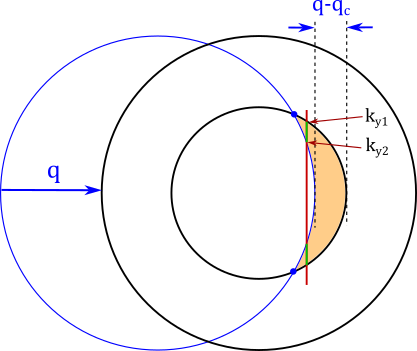}
    \caption{The orange shaded region is the area in $\vec k$  space where $f(\epsilon_{\vec k +\vec q}^+)-f(\epsilon_{\vec k}^-) \neq 0$. The damping rate $\Gamma$ is obtained by integrating over $\vec k$ inside this orange area. The delta function (see Eq.~\eqref{eq:damping delta function}) constrains the integral over $\vec k$ to a vertical (dark red)  line with a fixed $k_x = k_0$ for each given value of $\omega$ and $q$. The integral is therefore proportional to the length of the green line segment, which is written as $k_{y1}-k_{y2}$ in Eq.~\eqref{eq:damping Gamma result2}. The \addELH{horizontal distance between the vertical red line and the tips of the crescent shape (blue dots) is proportional to $\omega$---this can be read off from Eq.~\eqref{eq:damping k0}, remembering that the blue dots are the locations where $\epsilon_{{\vec k}+{\vec q}}^+ = \epsilon_{{\vec k}}^-$.} Therefore, for small $\omega$, the value of $\Gamma$, which is the length of the vertical line segment, is also proportional to $\omega$.}
    \label{fig:damping integral}
\end{figure}
\begin{align}
    \Gamma(\omega,q) &= \frac{\pi}{(2\pi)^2} \int_{\Omega_q} dk_x dk_y    \delta\left( k_x - k_0 \right) \cdot \frac{m_0}{q} \label{eq:damping Gamma result1}\\
    &= \Theta(q - q_c)\frac{m_0}{2 \pi q} (k_{y1}-k_{y2}) \label{eq:damping Gamma result2}\\
    & \propto  O(\omega)  \Theta(q - q_c), \label{eq:damping Gamma result3}
\end{align}
where $\Omega_q$ is a crescent shape area in $\vec k$-space (see Fig.~\ref{fig:damping integral}), which occurs when $q>q_c$. \addELH{The momenta $k_{y1}$ and $k_{y2}$ determine the length of the line segment being integrated over, as defined in Fig.~\ref{fig:damping integral}. For small frequencies $\omega \ll 2 \bar{m}$ (note that $\bar m$ is comparable to $\mu$ in our setting), the length of this line segment is proportional to $\omega$.}

Finally, \approve{we convert the damping rate to the Matsubara frequency expression for the transverse susceptibility, $\Pi^{+-}(i \nu)$} through the spectral representation:
\be
\Pi^{+-}(i\nu, {\vec q}) = \int_0^\Lambda \frac{d\omega}{2\pi} \frac{2\omega \Im \Pi^{+-}(\omega+i0^+, {\vec q})}{\nu^2+\omega^2} \propto |\nu|\Theta\left(|{\vec q}| - q_c\right) \int_0^{\Lambda/|\nu|} \frac{x^2 dx}{x^2+1} = F(q) + r|\nu|\Theta \left(|{\vec q}| - q_c\right)
\ee
In conclusion, the Landau damping of the magnon in 2D gives rise to a non-analytic leading term $\sim |\nu|$ in the transverse spin susceptibility. This result is in agreement with what is known in a 2D antiferromagnet
\cite{PhysRevB.51.14874}. \addELH{Plugging back into Eqs.~\eqref{eq:chi_1_result} and \eqref{eq:chi_2_result}, we find that $\chi_1$ is corrected by this non-analytic $O(|\nu|)$ term through Landau damping, whereas $\chi_2$ is unchanged.}

\approve{While this term  changes the frequency dependence of the spin susceptibility at small frequencies, it does not qualitatively change our analysis of pairing. To see this, we note that 
\begin{itemize}
    \item When entering the expression for the pairing interaction, the damping term's $O(|\nu|)$ correction always comes together with $\kappa \nu^2+Dq^2$; 
    \item The magnon modes that mediate the pairing interaction are at finite $q \sim k_0$.
\end{itemize}
Given these two facts, we see that the damping's $O(|\nu|)$ correction is subleading until it exceeds some finite value $\sim Dk_0^2$. Below, we estimate the frequency scale below which $|\nu|$ is negligible: The parameter $r$ is a function of Fermi-surface-related quantities (shape and curvature). Meanwhile, as a reminder, the scale of $q$ relevant for pairing is $\sim \sqrt{J/V} k_F$. As a result, by dimensional analysis, the frequency scale below which $|\nu|$ is negligible is $\sim O(\frac{J}{V}E_F)$. However, the frequency of the magnon that dominates the pairing is below $\frac{J}{V} E_F$ anyway. Therefore, we conclude that damping does not significantly alter the pairing interaction.
}

\end{widetext}

\end{document}